\documentclass[a4paper, onecolumn, 11pt, accepted=2025-01-25]{quantumarticle}
\pdfoutput=1
\usepackage[numbers, sort&compress]{natbib}
\usepackage[utf8]{inputenc}
\usepackage[english]{babel}
\usepackage[T1]{fontenc}
\usepackage{amsmath,amssymb,amsthm}
\usepackage{hyperref}
\usepackage{dsfont}
\usepackage{mathrsfs}
\usepackage{enumitem}
\usepackage{subfigure}

\usepackage{graphicx}
\usepackage{color}
\usepackage{upgreek}
\usepackage{float}
\usepackage{lipsum}
\usepackage{mathrsfs}
\usepackage{booktabs}
\usepackage{algorithm} 
\usepackage{algpseudocode}
\usepackage{acronym}

\newtheorem{definition}{Definition}

\algnewcommand\algorithmicinput{\textbf{Input:}}
\algnewcommand\algorithmicoutput{\textbf{Output:}}
\algnewcommand\Input{\item[\algorithmicinput]}
\algnewcommand\Output{\item[\algorithmicoutput]}

\newcommand{\ket}[1]{\ensuremath{\left| #1 \right\rangle}}
\newcommand{\bra}[1]{\ensuremath{\left\langle #1 \right|}}

\DeclareMathOperator*{\argmin}{\arg\!\min}
\newcommand{\Tr}[0]{\ensuremath{\text{Tr}}}

\newcommand{\kket}[1]{\ensuremath{\left| #1 \right\rangle\!\rangle}}

\AtBeginDocument{
\heavyrulewidth=.08em
\lightrulewidth=.05em
\cmidrulewidth=.03em
\belowrulesep=.65ex
\belowbottomsep=0pt
\aboverulesep=.4ex
\abovetopsep=0pt
\cmidrulesep=\doublerulesep
\cmidrulekern=.5em
\defaultaddspace=.5em
}
\begin{document}
 
\title{What can unitary sequences tell us about multi-time physics?}

\author{G. A. L. White}
\email{greg.al.white@gmail.com}
\affiliation{Dahlem Center for Complex Quantum Systems, Freie Universit\"at Berlin, 14195 Berlin, Germany}
\affiliation{School of Physics and Astronomy, Monash University, Clayton, VIC 3800, Australia}
\affiliation{School of Physics, University of Melbourne, Parkville, VIC 3010, Australia}

\author{F. A. Pollock}
\affiliation{School of Physics and Astronomy, Monash University, Clayton, VIC 3800, Australia}

\author{L. C. L. Hollenberg}
\affiliation{School of Physics, University of Melbourne, Parkville, VIC 3010, Australia}

\author{C. D. Hill}
\email{charles.hill1@unsw.edu.au}
\affiliation{Silicon Quantum Computing, The University of New South Wales, Sydney, New South Wales 2052, Australia}
\affiliation{School of Physics, University of Melbourne, Parkville, VIC 3010, Australia}
\affiliation{School of Mathematics and Statistics, University of Melbourne, Parkville, VIC, 3010, Australia}

\author{K. Modi}
\email{kavan\_modi@sutd.edu.sg}
\affiliation{Science, Mathematics and Technology Cluster, Singapore University of Technology and Design, \\8 Somapah Road, 487372 Singapore}
\affiliation{School of Physics and Astronomy, Monash University, Clayton, VIC 3800, Australia}

\begin{abstract}
Multi-time quantum processes are endowed with the same richness as multipartite states, including temporal entanglement and exotic causal structures. However, experimentally probing these rich phenomena leans heavily on fast and clean mid-circuit measurements, which are rarely available. We show here how surprisingly accessible these phenomena are in nascent quantum processors even when faced with substantially limited control.
We work within the limitation where only unitary control is allowed, followed by a terminating measurement. Within this setting, we first develop a witness for genuine multi-time entanglement, and then methods to bound (from top and bottom) multi-time entanglement, non-Markovianity, purity, entropy, and other correlative measures. Our tools are designed to be implemented on quantum information processors, which we proceed to demonstrate. Finally, we discuss the limitations of these methods by testing them across random multi-time processes. Conceptually, this broadens our understanding of the extent to which temporal correlations may be determined with only deterministic control.
Our techniques are pertinent to generic quantum stochastic dynamical processes, with a scope ranging across condensed matter physics, quantum biology, and in-depth diagnostics of NISQ-era quantum devices.
\end{abstract}

\maketitle
\tableofcontents

\section{Introduction}

Quantum information processors -- computing devices, sensors, and communicators -- have recently seen an explosion in both accessibility and interest~\cite{Bluvstein_2023, kim2023evidence, acharya2024quantumerrorcorrectionsurface, ryan2024high, pirandola2018advances}. Even more broadly, adjacent fields such as condensed matter physics and biology are becoming increasingly engaged with the direct effects of quantum mechanics in their respective areas~\cite{marx2021biology,mcfadden2018origins}. Crucially, dynamics of open quantum systems play a central common role in all of these areas, either as a source for complex noise, hidden signals, or even foundational physics. However, the ability to equip this nascent technology to study the dynamics of open quantum systems is still very much lacking. In this paper, we show that this is a problem that can be solved through methodology and conceptual breakthroughs, even when facing many limitations in controlling the hardware. Namely, we provide a new set of tools designed to understand the nature of temporal correlations in generic open quantum systems and, importantly, in the context of noise affecting quantum devices.

Microscopic correlations affect the macroscopic properties of many-body quantum systems, leading to exotic phases of matter. Discourse on this topic is typically dominated by spatial correlations exhibited at a single time, or where time is merely a parameter for investigating the evolution of spatial properties~\cite{horodecki2009quantum, lanyon2017efficient}. We contend that a single quantum system may have rich dynamics, especially when it interacts with its environment, with non-trivial multi-time (quantum) correlations~\cite{milz21,aloisio-complexity}. Such correlations must be accounted for to pave the pathway to fault-tolerant quantum computers. In other technologies, such as quantum sensors, these correlations could be studied in order to have a better qualitative and quantitative understanding of the environment. 
Comparatively little has been studied on the subject of this \emph{many-time} physics. 
One reason for the relative sparseness of attention has been because of the challenge in putting many-time correlations on equal footing as spatial ones. On the experimental front, however, the problem is even deeper. Certification of temporal correlations requires a set of ideal quantum instruments, such as a measure and feed-forward process. However, not only are current mid-circuit measurement processes unreliable, they are also relatively slow -- ranging from a few microseconds in superconducting devices~\cite{IBMQuantum}, to hundreds in ion traps~\cite{PRXQuantum.2.020343}, for example. This is problematic if the time to probe is commensurate with the timescale of the background dynamics under scrutiny.
This can make such characterisation at best unreliable, and at worst misleading. Additionally, many quantum experiments, such as quantum sensors, do not possess any mid-circuit measurement capabilities at all and instead operate with a sequence of control operations. However, this is not always the case and there are notable exceptions~\cite{Adams_2020, PhysRevX.13.041035}.

In this work, we address the question: how can we access the fantastic multi-time phenomena in quantum experiments and in near-term quantum hardware? By doing so, we open the door up to the realistic observation and study of temporal correlations on much the same footing as many-body states. 
This has timely applications relevant to quantum technologies: from model-independent environment determination, to novel sensing methods, diagnosis of temporally correlated noise, and avenues to quantum advantage via simulation of open systems -- as some examples. Moreover, it offers a unique insight into the nature and observation of multi-time quantum processes. To achieve this goal, we develop new techniques uniquely tailored for observables on temporal quantum states -- particularly for different control paradigms. 
We show that quantum computers, even in their current state, offer a timely opportunity to explore the richness of multi-time physics. While these machines are too noisy to manipulate into the large entangled registers required for interesting quantum algorithms, they are riddled with complex noise~\cite{White-NM-2020,White-MLPT}, which we can access with high-fidelity single qubit unitary controls. Thus, we show that there is scope to explore interesting physics even with relatively small devices -- a single qubit probed across many times, for example, can exhibit many of the same properties as large quantum states.

We begin with a brief introduction to quantum stochastic processes and multi-time correlations in Sec.~\ref{sec:background} in the language of process tensors. We then pose the question `what can we learn about a quantum process when we can only exercise limited control?' Addressing this question is the core theme of our work, and has applications to our understanding of temporal correlations; the capabilities of quantum sensors and metrology; and the diagnosis of complex noise in near-term intermediate-scale quantum (NISQ) devices. In Sec.~\ref{sec:temp-entanglement}, we show how using only unitary control, temporal entanglement can be witnessed. Then in Sec. \ref{sec:bounding}, we bound a whole host of non-Markovian correlation measures, again working within the paradigm of limited control. We benchmark our techniques in Sec. \ref{sec:bm} with random processes. Our conclusions are presented in Sec. \ref{sec:dis}.

\section{Background: Quantum Stochastic Processes}
\label{sec:background}

In any given experiment, there is a distinction between controllable operations implemented by the experimenter on a $d-$dimensional system, and uncontrollable stochastic dynamics due to system-environment ($SE$) interactions. A quantum stochastic process accounts for all quantum correlations across a sequence of time-ordered events at time times $\mathbf{T}_k:=\{t_0,\cdots, t_k\}$. For instance, we may desire to know the nature of correlated noise on a quantum computer, where the allowed control is a sequence of gates applied at times $\mathbf{T}_k$ followed by a terminating measurement. A quantum sensor acquires a signal from its environment and by applying appropriate control we can learn the nature of the environment from the temporal correlations in the signal. Recent developments have led to a comprehensive theory for describing and characterising temporal quantum correlations. Here, we continue on that journey to characterise complex features of multi-time correlations under realistic control limitations.

{Aside from adhering to real-world practicalities, the restriction to unitary sequences is conceptually interesting for the study of quantum stochastic processes. In the following, let us consider the case where all measurements are projective. Typically, when considering correlations of any type, one might think of three systems $A$, $B$, and $C$ owned by Alice, Bob, and Charlie separated in space, time, or both. To observe spatial correlations, the players can coordinate measurement bases and compare outcomes. To measure correlations in the temporal setting, or non-Markovian correlations, the players can perform the same experiment (in a time ordered fashion) with an extra step that between $t_A$ and $t_B$, the system is deterministically reset to some state $\rho_0$. After evolution to time $t_B$, Bob measures the state. And then similarly onto Charlie. This may be summarised by the following:
\begin{equation}
	\label{eq:entanglement-description}
	\stackrel{\text{Alice measures}}{\overbrace{\rho_A}}
 \xrightarrow{\text{Reset}} 
 \ \rho_0 \ \xrightarrow{\text{Evolution}} \stackrel{\text{Bob measures}}{\overbrace{\rho_B}}
 \xrightarrow{\text{Reset}} 
 \ \rho_0 \ 
 \xrightarrow{\text{Evolution}} \stackrel{\text{Charlie measures}}{\overbrace{\rho_C}}.
\end{equation}
Above, Alice observes an outcome $a$, Bob observes an outcome $b$, and Charlie observes $c$, from which temporal correlations can be inferred.

It can be shown that any non-classical correlations detected in this experiment will signify quantum non-Markovianity~\cite{Giarmatzi2021, milz21}. Because a causal break -- the ability to reset the system to state $\rho_0$ -- is applied, the correlations can only persist from an external environment, which we shall call $E$. Suppose $\rho_A$ is the marginal of some larger correlated (entangled) state $\rho_{AE}$. Then a measurement on $A$, with outcome $a$, will project $E$ onto a state $\rho_{E|a}$ by virtue of their shared correlation. Then, suppose the evolution $\mathcal{E}_{t_A\rightarrow t_B}$ is also entangling. When Bob measures, the collapse at $B$ will depend on the collapse of the environment at time $t_A$, and hence on the outcome $a$ of Alice. For Charlie both of the previous collapses will matter. If the correlations are superclassical, then the environment has distributed quantum information across times (temporal entanglement).
As hinted at in Ref.~\cite{aloisio-complexity}, and as we shall see in this work, it does not suffice for $\mathcal{E}_{t_A\rightarrow t_B}$ to be merely entangling, it must be of a non-diagonal form to generate temporal entanglement. The players can now coordinate bases and compare measurement statistics to determine whether the environment was sufficiently complex to share entanglement across time. One can also generalise these scenarios to genuine multipartite entanglement across $k$-step processes~\cite{milz21}.}

{Let us now consider an alternate situation in line with the titular description. Alice and Bob are no longer permitted to measure their states, and the system is no longer reset. Instead, Alice and Bob may apply any unitary operation they like to the state. Only Charlie now measures the state after both of these actions. This can be summarised as
\begin{equation}
	\label{eq:unitary-description}
	\stackrel{\text{Alice applies } U_A}{\overbrace{\rho_A}} \xrightarrow{\text{Evolution}} \stackrel{\text{Bob applies }U_B}{\overbrace{\rho_B}}\xrightarrow{\text{Evolution}} \stackrel{\text{Charlie measures}}{\overbrace{\rho_C}}.
\end{equation}
The question is then: from knowledge of any $U_A$ and $U_B$ that Alice and Bob may apply (and his measurement statistics), what can Charlie say about the ability of Alice and Bob to measure Bell-like correlations, as in the situation described by Eq.~\eqref{eq:entanglement-description}? On the surface, it is not obvious at all that Charlie can glean any information. After all, Alice and Bob are performing completely deterministic operations--they have not collected any measurement statistics. However, as we shall see, a surprising amount can be learned about temporal correlations from this setup.}

We begin with a brief review of the theory of quantum stochastic process. Here, the central objects include completely positive operations, process tensors, Choi representation, instruments, testers, etc. This first requires the background of quantum channels, which embody two-time correlations. Once we re-state the familiar concepts there, we will generalise them for the multi-time setting.

\subsection{Quantum channels}
\label{app:channels}

\begin{figure}[t!]
    \centering
    \includegraphics[width=\linewidth]{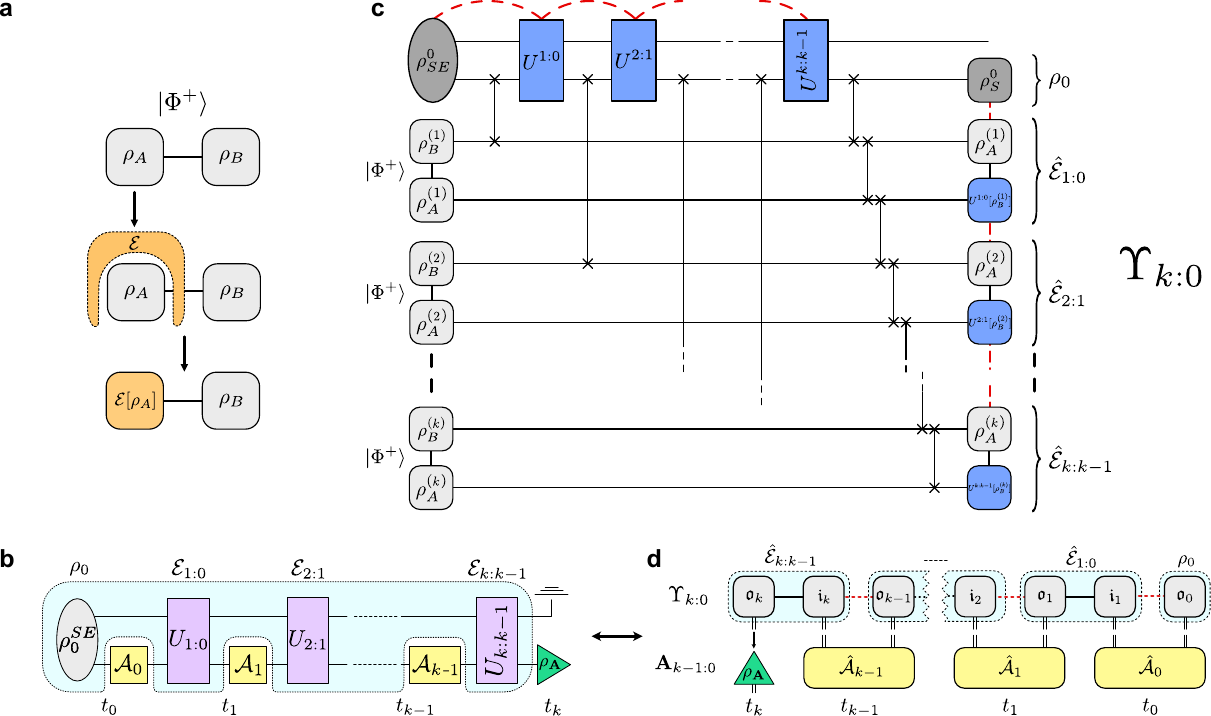}
    \caption{Circuit diagrams of the conventional and generalised Choi-Jamio\l kowski isomorphism. \textbf{a} Two-time processes are represented by quantum channels; their Choi state is given by the channel acting on one half of a Bell pair. 
    \textbf{b} Many-time processes are represented by process tensors, a multilinear mapping from operations $\mathcal{A}_i$ at different times to a final state conditioned on those operations.
    \textbf{c} A depiction of the realisation of the Choi-Jamio\l kowski isomorphism for process tensors. The Choi state is given for a $k$-step process by swapping in one-half of $k$ Bell pairs at different times. The non-Markovian correlations are mapped onto spatial correlations in the Choi state. The marginals of the process tensor are quantum channels, as well as the average initial state. Many of the exotic features of many-body physics can be linked to processes, thus generating an equivalent many-time physics. The space of processes is not isomorphic to all states, but rather states that obey a set of causality conditions.
    \textbf{d} A $k$-step process tensor Choi state. This object is equivalent to the final state of the circuit diagram in \textbf{c} and represents a sequence of possibly correlated CPTP maps between different points in time, plus the affect due to the initial state. Output legs $\mathfrak{o}_l$ are mapped by a control operation $\mathcal{A}_l$ to the next input leg $\mathfrak{i}_{l+1}$. Note the left-to-right time convention for circuit diagrams in \textbf{b}, contrasting right-to-left induced by matrix/tensor contraction in part \textbf{d}.
    }
    \label{fig:choi_reps}
\end{figure}

A completely positive (CP) channel is the mapping from bounded linear operators, $\mathscr{B}(\mathcal{H}_{\text{in}})\rightarrow \mathscr{B}(\mathcal{H}_{\text{out}})$ to bipartite quantum states on the respective spaces, $\mathscr{B}(\mathcal{H}_{\text{out}})\otimes \mathscr{B}(\mathcal{H}_{\text{in}})$.
\begin{equation}
\label{eq:cpmap}
    \mathcal{E}[\rho_{\text{in}}] = \text{tr}_E [u \ \rho_{\text{in}} \otimes \rho^{E} u^\dag] = \rho_{\text{out}}.
\end{equation}
Here, the potentially non-unitary process between the input and the output state is a result of unitary interaction with an environment $E$. 
Such a channel is often represented in terms of its Choi state $\hat{\mathcal{E}}$ by the projection
\begin{equation}
\label{eq:choi-act}
    \mathcal{E}[\rho_{\text{in}}] = \text{Tr}_{\text{in}}\left[(\mathbb{I}_{\text{out}}\otimes \rho_{\text{in}}^{\text{T}})\hat{\mathcal{E}}\right].
\end{equation}
Explicitly, the Choi state $\hat{\mathcal{E}}$ of a channel $\mathcal{E}$ on a $d-$dimensional system is formed by the action of $\mathcal{E}$ on one half of an unnormalised maximally entangled state $\ket{\Phi^+} = \sum_{i=1}^d \ket{ii}$:
\begin{equation}
    \hat{\mathcal{E}} := (\mathcal{E}\otimes \mathcal{I})\left[|\Phi^+\rangle\!\langle\Phi^+|\right] = \sum_{i,j=1}^d \mathcal{E}\left[|i\rangle\!\langle j|\right]\otimes |i\rangle\!\langle j|,
\end{equation}
where $\mathcal{I}$ is the identity map. This is depicted in Figure~\ref{fig:choi_reps}a, along with its many-time generalisation, which we introduce in the next subsection. 

\subsection{Process tensors}\label{sec:PT}
Quantum channels account for two-time correlations, between the input and output states. As stated above, in general, a quantum experiment may have a sequence of control operations applied to the system at times $\mathbf{T}_k:=\{t_0,\cdots, t_k\}$. Such a process is represented as a multilinear map from a sequence of controllable operations on the system to a final state density matrix. Here, the map, known as a process tensor~\cite{Pollock2018a, 1367-2630-18-6-063032}, represents all of the uncontrollable dynamics of the process, depicted as the background in Figure~\ref{fig:choi_reps}b.

Process tensors formally generalise quantum channels to many-time processes.
Namely, we are interested in multi-time quantum correlations to generalise the two-time correlations embodied by quantum channels. To be precise, we consider the situation where a $k$-step process is driven by a sequence $\mathbf{A}_{k-1:0}$ of control operations, each represented mathematically by CP maps: $\mathbf{A}_{k-1:0} := \{\mathcal{A}_0, \mathcal{A}_1, \cdots, \mathcal{A}_{k-1}\}$, after which we obtain a final state $\rho_k(\mathbf{A}_{k-1:0})$ conditioned on this choice of interventions. 
These controlled dynamics have the form:
\begin{equation}\label{eq:multiproc}
        \rho_k\left(\textbf{A}_{k-1:0}\right) = \text{tr}_E [U_{k:k-1} \, \mathcal{A}_{k-1} \cdots \, U_{1:0} \, \mathcal{A}_{0} (\rho^{SE}_0)],
\end{equation}
where $U_{k:k-1}(\cdot) = u_{k:k-1} (\cdot) u_{k:k-1}^\dag$ and $\mathcal{A}_{j}$ is the CP map applied at time $j$.

Eq.~\eqref{eq:multiproc} is the many-time generalisation of Eq.~\eqref{eq:cpmap}, where $\rho_{\text{in}}$ is now replaced by a sequence of control operations $\textbf{A}_{k-1:0}$. The process tensor $\mathcal{T}_{k:0}$ is defined as the mapping from past controls $\mathbf{A}_{k-1:0}$ to future states $\rho_k\left(\textbf{A}_{k-1:0}\right)$:
\begin{equation}
\label{eq:PT}
\mathcal{T}_{k:0}\left[\mathbf{A}_{k-1:0}\right] = \rho_k(\mathbf{A}_{k-1:0}).
\end{equation}

It is usually convenient to work with the Choi state of the process tensor. A generalisation of the \textit{Choi-Jamio{\l}kowski Isomorphism} (CJI) allows this mapping to be represented as a many-body quantum state. The circuit view of this CJI is depicted in Figure~\ref{fig:choi_reps}c, along with the usual CJI for quantum channels. In this picture, at each time, one half of a fresh Bell pair is swapped to interact with the environment. The mixed state at the end, $\Upsilon_{k:0}$ is the process tensor Choi state, and the control operations Choi states are denoted as $\hat{\mathcal{A}_j}$. This can be used to produce the action of the process tensor on any controlled sequence of operations (consistent with the time steps) by projection:
\begin{equation}\label{eq:PToutput}
    \rho_k(\mathbf{A}_{k-1:0}) \!=\! \text{Tr}_{\overline{\mathfrak{o}}_k} \! \left[ \left(\mathbb{I}_{\mathfrak{o}_k}\otimes \hat{\mathcal{A}}_{k-1}\otimes \cdots \hat{\mathcal{A}}_0 \right)^\text{T} \Upsilon_{k:0} \right],
\end{equation}
where $\overline{\mathfrak{o}}_k$ is every index except $\mathfrak{o}_k$, see Figure~\ref{fig:choi_reps}d. This equation is inclusive of all intermediate $SE$ dynamics as well as any initial correlations, and illustrates how sequences of operations constitute observables of the process tensor. This is a generalisation of Eq.~\eqref{eq:choi-act}. We emphasise our conventions on time throughout this work: in circuit diagrams, we have time running from left to right in accordance with the literature. In matrix and tensor representations, however, we have time running from right to left, in accordance with the standards of matrix multiplication. Although the tension between these two is unfortunate, we believe it to simplify matters in practice.

\subsection{Properties of quantum processes}\label{sec:props}

The Choi state is an operator on multipartite Hilbert spaces as
\begin{equation}\label{eq:processtensor}
\begin{split}
&    \Upsilon_{k:0} \in \mathscr{B}(\mathcal{H}_{\mathfrak{o}_k} \otimes \mathcal{H}_{\mathfrak{i}_{k-1}} \otimes \mathcal{H}_{\mathfrak{o}_{k-2}} \otimes \ldots \otimes \mathcal{H}_{\mathfrak{i}_{1}} \otimes \mathcal{H}_{\mathfrak{o}_0})\\
\text{with} \quad & \Tr[\Upsilon_{k:0}]=1, \ \Upsilon_{k:0}=\Upsilon_{k:0}^\dag, \ \Upsilon_{k:0} \ge 0.
\end{split}
\end{equation}
Each time has an associated output ($\mathfrak{o}$) and input ($\mathfrak{i}$) leg from the process. The Choi states in Eq.~\eqref{eq:nmpt} possesses the same features of states of many-body physics, but deterministic processes have causal constraints which mean that they map to only a subset of quantum states. Thus, we have the temporal parallel with many-time observables and many-time states. The many-time states possess the same richness of correlations, including being separable, discordant, entangled, and even different temporal phases. We dub this as \textit{many-time physics} because temporal quantum correlations are endowed with the same richness as spatial quantum correlations. Albeit, the interpretations of temporal entanglement are not always identical~\cite{milz21}. The generation of temporal correlations across multiple times has a different physical origin, however. They emerge when the system interacts strongly with a complex environment. Information can travel into a (quantum or classical) bath and return at a later point in time to interact with the system, distributing correlations~\cite{Milz2021PRXQ, rivas-NM-review}. The properties of this non-Markovian memory depend on both the interaction and the bath coherence.

All non-Markovian quantum stochastic processes are fully describable under the process tensor framework~\cite{Pollock2018a, Milz2021PRXQ}. More general processes with, e.g., indefinite causal orders, can employ the process matrix formalism~\cite{Shrapnel_2018}. Importantly, given the process tensor, one can compute many of its properties, e.g. different notions of non-Markovianity (memory length strength), temporal entanglement to certify the quantum nature of the environment, and the purity and entropies of the process to gauge the level of noisiness of the process. We briefly outline the structural properties of the process tensor below and how might we detect non-Markovianity and temporal entanglement.

If the dynamical process is non-Markovian, then the system-environment interactions will distribute temporal correlations as spatial correlations between different legs of the process tensor. These correlations may then be probed using any number of established quantum or classical many-body tools. In Figure~\ref{fig:choi_reps}b we have depicted these correlations in red. In full generality, the process tensor can be expanded in terms of a family of CP maps as
\begin{equation}\label{eq:nmpt}
    \Upsilon_{k:0}^{(\text{non-Markov})} = \sum_{\sigma} \alpha_\sigma \ \hat{\mathcal{E}}_{k:k-1}^{\sigma}\otimes \hat{\mathcal{E}}_{k-1:k-2}^{\sigma}\otimes\cdots\otimes \hat{\mathcal{E}}_{1:0}^\sigma \otimes\rho_{0}^{\sigma},
\end{equation}
where $\{\hat{\mathcal{E}}_{y:x}^\sigma\}$ are the Choi states of CP maps from time $x$ to $y$. The index $\sigma$ and the amplitudes $\alpha_\sigma$ account for correlations in time. For classical non-Markovian processes, each $\hat{\mathcal{E}}^\sigma$ is a completely positive and trace-preserving (CPTP) channel and $\alpha_\sigma \ge 0$.\footnote{If $\mathcal{E}$ are CP but not TP then one may require a quantum environment~\cite{taranto2023characterising}.}

The correlations between consecutive legs, $\mathfrak{i}_j$ and $\mathfrak{o}_j$ (see Fig.~\ref{fig:choi_reps}d), are expected to be very high for nearly-unitary processes and do not constitute a measure of non-Markovianity. In this vein, it can be shown that a process is Markovian if and only if it can be expressed in the form~\cite{Pollock2018}:
\begin{equation}
\Upsilon_{k:0}^{(\text{Markov})} = \hat{\mathcal{E}}_{k:k-1}\otimes \hat{\mathcal{E}}_{k-1:k-2}\otimes\cdots\otimes \hat{\mathcal{E}}_{1:0}\otimes\rho_0,
\end{equation}
where all $\hat{\mathcal{E}}$ are required to be CPTP. To operationally detect non-Markovian correlations in a process, we need to test if the future process is correlated with the past process. This can be done using Eq.~\eqref{eq:PToutput} and a \textit{causal break}, i.e., looking inequalities of the following form
\begin{equation}\label{eq:NMC}
\frac{\mbox{Tr}[(\mathbf{A}^{(y)}_{\mathfrak{o}_{k} \cdots \mathfrak{i}_{j+1} } \otimes \mathbf{A}^{(x)}_{\mathfrak{o}_j \cdots \mathfrak{o}_{0}} )^\text{T} \Upsilon_{k:0}]}{\mbox{Tr}[(\mathbf{A}^{(x)}_{\mathfrak{o}_j : \mathfrak{o}_{0}} )^\text{T} \Upsilon_{k:0}]}
\!\ne\!
\frac{\mbox{Tr}[(\mathbf{A}^{(y)}_{\mathfrak{o}_{k} \cdots \mathfrak{i}_{j+1} } \otimes \mathbf{A}^{(x')}_{\mathfrak{o}_j \cdots \mathfrak{o}_{0}} )^\text{T} \Upsilon_{k:0}]}{\mbox{Tr}[(\mathbf{A}^{(x')}_{\mathfrak{o}_j : \mathfrak{o}_{0}} )^\text{T} \Upsilon_{k:0}]}.
\end{equation}
The above equation asks whether the probabilities of observing an outcome $y$ in the future $(\mathfrak{o}_{k} \cdots \mathfrak{i}_{j+1})$ process differ, given that we observed two distinct outcomes $x$ and $x'$ in the past at $(\mathfrak{o}_{j} \cdots \mathfrak{o}_{0})$? The denominators are the probability of observing $x,x'$, which we do not care about. Importantly, we must break up the instrument at $j$ into a past and a future component -- across $\mathfrak{o}_j$ and $\mathfrak{i}_{j+1}$ -- this is the aforementioned causal break. It ensures that the future depends on the past if and only if the process is Markovian. This is because, when the instrument at $j$ is broken no information passes from the past to the future via the system.\footnote{This condition can be made stronger by multi-time conditionings, i.e. we may think of $y,x,x'$ to be a sequence of outcomes.} Note, that to test the above inequality, at least two measurements must be made in a single experiment, one at $\mathfrak{o}_k$ and another at $\mathfrak{o}_j$.

It is important to note that although the set of process tensors is isomorphic to a subset of quantum states, processes are much more restricted in terms of the structure they can possess due to causality constraints. Causality, that the future cannot affect the past, appears in the form of a containment property in process tensors, that averaging over measurements on an output leg produces the exact process up until that point: 
\begin{equation}
    \label{eq:causality}
    \text{Tr}_{\mathfrak{o}_k}[\Upsilon_{k:0}] =\mathbb{I}_{\mathfrak{i}_j}\otimes\Upsilon_{k-1:0}.
\end{equation}
This provides a time ordering to the structure of correlations in processes. One example consequence is that rank one processes must be purely Markovian when the system dimension is constant -- i.e. any many-time structure must be mediated by the surrounding environment. The causal requirement and the mechanism by which correlations are accessed are the primary difference between states and processes, and the subject of our exploration in this work. As we will discuss, process observables are non-trivial to deal with, and have different structures to those of state observables -- for example, they can be deterministic at different times. We lastly remark that as an mapping (i.e., for valid action on quantum operations), the trace of the process tensor Choi matrix must be appropriately set to $d^{k}$ (where $d$ is the system dimension). However, in computing properties of the `Choi state', this is normalised to unit trace. We will make the distinction clear through context.

\subsection{Process tensor tomography}
\label{methods:PTT}

Due to the state-like structure of the process tensor, it's straightforward to apply the usual process tomography tools for reconstructing a process. Said differently, the process tensor framework is operationally built to experimentally access many-time quantum correlations. Recently, we have introduced process tensor tomography (PTT) to facilitate the characterisation of non-Markovianity on real devices~\cite{White-NM-2020, White-MLPT}. The rough idea here is to sample from a quantum process and use the relative frequencies of the observation to infer the process, much like any other tomography procedure.

We start with measuring the final state of the system, given in Eq.~\eqref{eq:PToutput},  with measurement operator $\{\Pi_x\}$ at time $k$. The output distribution, conditioned on these operations and measurement apparatus, is then given by the spatiotemporal generalisation of Born's rule~\cite{Shrapnel_2018}:
\begin{equation}
    \label{eq:PT_action}
    p_{x|\mathbf{A}_{k-1:0}} = \text{Tr}\left[\Upsilon_{k:0}(\Pi_x \otimes  \mathbf{A}_{k-1:0}^{\text{T}})\right].
\end{equation}
This generalises joint probability distributions of a classical stochastic process to the quantum regime~\cite{Milz2020}. The key insight to Eq.~\eqref{eq:PT_action} is that operations on a system at different points in time constitute many-time observables on the process tensor Choi state. These may be deterministically applied -- such as a unitary operation -- or stochastic, such as a measurement and feed-forward or more general quantum instruments. Note that many-time processes suffer from the same dimensionality curse as states and multipartite classical distributions. Here, with the number of timesteps, the number of histories to account for grows exponentially.

Although Eq.~\eqref{eq:PT_action} can in principle be inverted to determine $\Upsilon_{k:0}$, in practice sampling statistics and instrument error will lead to an estimate which is neither positive nor causal. To circumvent this, and construct an optimal model, maximum-likelihood estimation (MLE) PTT seeks a positive and causal process which maximises consistency with the data, as described in Ref.~\cite{White-MLPT}. We briefly outline MLE-PTT now. The first step is to pick a basis for the terminating measurement $\{\Pi_x\}$ and the quantum operations $ \{ \mathcal{B}_j^{\mu_j}\}$. If $\text{span}(\{\Pi_x\}) = \mathscr{B}(\mathcal{H}_{\mathfrak{o}_k})$ and $\text{span}(\{ \mathcal{B}_j^{\mu_j}\}) = \mathscr{B}(\mathcal{H}_{\mathfrak{i}_{j+1}}\otimes \mathcal{H}_{\mathfrak{o}_j})$ then the control is said to be \emph{informationally complete} (IC) and can hence be used to determine the entire process. If we can estimate the corresponding outcome probabilities, $p_{i,\vec{\mu}}$, then, via a set of linear equations, uniquely fixes $\Upsilon_{k:0}$. That is, applying Eq.~\eqref{eq:PT_action} to the basis elements yields
\begin{equation}
    p_{i,\vec{\mu}} = \text{Tr}\left[ \Upsilon_{k:0} (\Pi_i\otimes \mathcal{B}_{k-1}^{\mu_{k-1}}\otimes \cdots \otimes \mathcal{B}_0^{\mu_0})^\text{T} \right].
\end{equation}
Alternatively, if the probabilities are known only for some subspace of operations (in this work, we are frequently restricted to the space of unitary operations), then the process tensor will be uniquely fixed on that subspace. 

However, in reality, the measured outcomes $m_{i,\vec{\mu}}$ are noisy estimates of the `true' probabilities, and there is often no physical process tensor that completely matches the data. For this reason -- as is common -- we treat the estimation of a process as an optimisation problem where a model for the process is fit to the data. Specifically, a unique $\Upsilon_{k:0}$ is found by minimising the log-likelihood
\begin{equation}\label{eq:mle}
    f(\Upsilon_{k:0}) = \sum_{i,\vec{\mu}}-m_{i,\vec{\mu}} \ln p_{i,\vec{\mu}} \ ,
\end{equation}
where $\Upsilon_{k:0}$ is a positive matrix obeying a set of causality conditions given in Eq.~\eqref{eq:causality}. If we denote the cone of $n\times n$ positive semi-definite (PSD) matrices by $\mathcal{S}_n^+$ and the affine space laid out by causality requires by $\mathcal{V}_{\text{c}}$ then $\Upsilon_{k:0}\in\mathcal{S}_n^+\cap \mathcal{V}_{\text{c}}$. 
This optimisation is carried out using a projected gradient descent algorithm, where at each step the model is iterated in a direction that decreases the log-likelihood, followed by an orthogonal projection onto $\mathcal{S}_n^+\cap \mathcal{V}_{\text{c}}$. The optimisation has a guaranteed convergence, and the model can be tested by comparing its predictions to experiments. In general, MLE not only produces a physical model but an extremely robust one, with prediction fidelities in the realm of 99.9\%. More details on this estimation can be found in Ref.~\cite{White-MLPT}.

\subsection{Partitioning observable constraints by available control}\label{ssec:partitioning-control}
We will now analyse the measurement primitives offered by different control operations. For this, we will consider a single qubit system in the Pauli basis, but our results generalise readily to $d$-dimensional systems.
For most currently available quantum computers, one-qubit gates are much faster than two-qubit gates, which are much faster than measurements. The large timescale required for measurements means that mid-circuit measurements are usually not available. This usually does not pose many problems as one usually wants to apply a sequence of unitary gates and then make a terminating measurement. The limitation to only unitary control means that it is not possible to uniquely determine $\Upsilon_{k:0}$, which requires an IC control basis.

Yet, non-Markovian quantum tomography under limited control is possible and has been studied both theoretically~\cite{kuah,PT-limited-control} and experimentally~\cite{White-NM-2020,White-MLPT}. The reconstructed process, under limited control, is often referred to as a \textit{restricted process tensor}. `Restricted', here, is a somewhat nebulous term that refers to any process tensor characterised with a tomographically incomplete set of operations. In our context, we will refer exclusively to the restriction to unitary control. Whereas the dimension of a full process tensor is $\mathcal{O}(d^{4k})$, a restricted process tensor is $\mathcal{O}((d^4 - 2d^2)^{k})$-dimensional. While a restricted process tensor is perfectly capable of characterising the dynamics, it lacks certain desirable properties of the full process tensor. For instance, a restricted process tensor is not a complete matrix and does not have well-defined positivity conditions~\cite{Milz2017}. In other words, it is not a density matrix and cannot be used to compute temporal correlations straightforwardly. This includes temporal entanglement, non-Markovianity, and even the purity of the process. Plainly, its function is as a multi-linear mapping, but cannot take advantage of the nice properties of the CJI.

This raises the question \textit{`Can we deduce any properties of multi-time processes using only limited control?'} In this paper, we will show that indeed we can infer a great deal about a quantum process by processing only the restricted process tensor. While we confine ourselves to the limitation of unitary control and terminating measurements, the underlying philosophy of this paper can be applied to any limitation on control. Of course, each limitation will require a separate analysis. Many of the tools developed here will aid in such analysis. 

To understand what we can and cannot do under the unitary control limitation, we begin by looking at the detailed structure of unitary observables. To start, it is useful to express this problem using the Pauli basis, $\mathbf{P}^n :=\{\mathbb{I},X,Y,Z\}^{\otimes n}$. This way, the process tensor can be expressed as: $\Upsilon_{k:0} = \sum_{i=1}^{4^{2k+1}} \Tr[\mathsf{P}_i\Upsilon_{k:0}]\mathsf{P}_i$ with $\mathsf{P}_i \in \mathbf{P}^n$. Recalling that we label the marginal indices of $\Upsilon_{k:0}$ by interleaving $\mathfrak{o}_j$ and $\mathfrak{i}_j$ subsystems
\begin{equation}
	\Upsilon_{k:0} = \sum_{\vec{\mu}
 } \Tr[P^{\mu_{\mathfrak{o}_k}}\otimes P^{\mu_{\mathfrak{i}_k}} \otimes \cdots \otimes P^{\mu_{\mathfrak{o}_0}} \Upsilon_{k:0}] \ P^{\mu_{\mathfrak{o}_k}}\otimes P^{\mu_{\mathfrak{i}_k}} \otimes \cdots \otimes P^{\mu_{\mathfrak{o}_0}} ,
\end{equation}
where $P^{\mu_{\mathfrak{i}_j}}, P^{\mu_{\mathfrak{o}_j}}\in \mathbf{P}^1$. For brevity, we will write the above equation as an inner product $\langle\!\langle P^{\mu_{\mathfrak{o}_k}}\otimes P^{\mu_{\mathfrak{i}_k}} \otimes \cdots \otimes P^{\mu_{\mathfrak{o}_0}} | \Upsilon_{k:0} \rangle\!\rangle$. The causality condition in Eq.~\eqref{eq:causality} would pose restrictions on some of the terms in the above equation, e.g.
\begin{equation}
    \label{eq:causalityPauli}
    \langle\!\langle \mathbb{I}^{\mathfrak{o}_k} \otimes P^{\mu_{\mathfrak{i}_k}} \otimes \mathsf{P}_{k-2} | \Upsilon_{k:0}\rangle\!\rangle = 0 \quad \forall \quad P^{\mu_{\mathfrak{i}_k}} \ne \mathbb{I}^{\mathfrak{i}_k}.
\end{equation}
Further conditions on $\mathsf{P}_{k-2}$ follow by the same logic. We denote the set of Pauli strings instruments that are vanishing as above as $\mathfrak{C}$.\footnote{For the identity Pauli string instrument $\bigotimes\mathbb{I}$, we have $\langle\!\langle \bigotimes\mathbb{I}\|\Upsilon_{k:0}\rangle\!\rangle = 1$, which follows from normalisation of the Choi state given in Eq.~\eqref{eq:processtensor}.}

\begin{figure}
	\centering
	\includegraphics[width=0.25\linewidth]{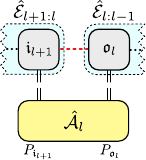}
	\caption[The projection of a process tensor onto some local control operation]{The projection of a process tensor onto some local control operation $\mathcal{A}_l$. We see that this evaluates expectation values on $\mathfrak{o}_l$ and $\mathfrak{i}_{l+1}$, and can be partitioned into Pauli tensor products on each of these legs.}
	\label{fig:pauli-expectations}
\end{figure}

Given that control instruments can also be expressed in a (restricted) Pauli basis, the non-zero coefficients will tell us what we can learn about the process. This notion is expressed in Figure~\ref{fig:pauli-expectations}, where the legs of an operation $\hat{\mathcal{A}}_l$ are projected onto the marginals of a process tensor Choi state. 
In the spatial setting, local and IC observables are readily available in any typical experimental setting. Temporally, however, learning observables of a quantum stochastic process can be more restricted in terms of the available instruments. In particular, the ability to learn mid-circuit information about a quantum state is often a long, invasive, and technologically challenging operation.

For gate-based quantum devices, the first obvious port-of-call are unitary channels
\begin{equation}
\begin{split}
\hat{\mathcal{A}}_{\text{u}} &= (u\otimes \mathbb{I})\cdot |\Phi^+\rangle\!\langle \Phi^+| \cdot(u^\dagger \otimes \mathbb{I})\\
    &= \sum_{i,j}u(|i\rangle\!\langle j|)u^\dagger \otimes |i\rangle\!\langle j|
\end{split}
\end{equation}
for unitary $u$. This has only non-trivial Pauli coefficients for $P_{\mathfrak{o}}$ and $P_{\mathfrak{i}}$ both traceless or both full trace, as follows:
\begin{equation}
	\begin{split}
\Tr[{P}_\mathfrak{o}\otimes P_{\mathfrak{i}} \hat{\mathcal{A}}_{\text{u}}] &= \sum_{ijkl}\langle k|P_\mathfrak{o} u|i\rangle\!\langle j|u^\dagger|k\rangle \otimes \langle l|P_{\mathfrak{i}}|i\rangle\!\langle j|l\rangle\\
	&= \Tr[u P_{\mathfrak{i}}^{\text{T}}u^\dagger P_{\mathfrak{o}}].
	\end{split}
\end{equation}
We see from this that $\langle {P}_{\mathfrak{o}}\mathbb{I}_{\mathfrak{i}}\rangle = \langle \mathbb{I}_{\mathfrak{o}}{P}_{\mathfrak{i}}\rangle = 0$ by the unitarity of $u$ when $P$ is traceless.

A physical way to interpret this is that unitary maps provide no information about the process marginals; they do not allow one to read out information about the state at $\mathfrak{o}_{l}$, nor do they allow one to know about the state entering the process at $\mathfrak{i}_{l+1}$. Instead, a sequence of unitary operations is equivalent to teleportation of the state through the process and applying operations to it along the way.
Thus, when restricted only to unitary operations (and a final measurement) on a single qubit, the observables $\mathcal{O}$ on $\Upsilon_{k:0}$ -- to which we have access -- take the form
\begin{equation}
	\label{eq:restr-obs}
	\begin{split}
	&	\mathcal{O} = \Pi_x\otimes \sum_{i,\vec{\mu}}\alpha_{\vec{\mu}}  \bigotimes_{j=0}^{k-1} 
 P^{\mu_{\mathfrak{i}_{j+1}}\mu_{\mathfrak{o}_j}}, \qquad \text{where}\\
		&\alpha_{\vec{\mu}}\in \mathbb{R},\quad
P^{\mu_{\mathfrak{i}_{j+1}}\mu_{\mathfrak{o}_j}} \in \{\mathbb{I\otimes I}\} \cup \{X,Y,Z\}\otimes \{X,Y,Z\}.
	\end{split}
\end{equation}
Here, $\{\Pi_x\}$ is IC, but the $P^{\mu_{\mathfrak{i}_{j+1}}\mu_{\mathfrak{o}_j}}$ do not include a single local $\mathbb{I}$ term on either $\mathfrak{i}_{j+1}$ or $\mathfrak{o}_{j}$ alone. Such terms belong to non-unital and trace-decreasing operations, with the former to prepare coherent states on $\mathfrak{i}_{j+1}$ and the latter to be able to measure the current state at $\mathfrak{o}_{j}$. These are precisely the types of operations required to certify and quantify non-Markovianity, as described above in Eq.~\eqref{eq:NMC}. This observation is also true for generic unital (or bistochastic) quantum operations. Although these cannot be expressed as a convex combination of unitary operations, they do lie within the linear span of unitaries~\cite{mendl2009unital}.

\begin{figure}[t!]
	\centering
	\includegraphics[width=0.8\linewidth]{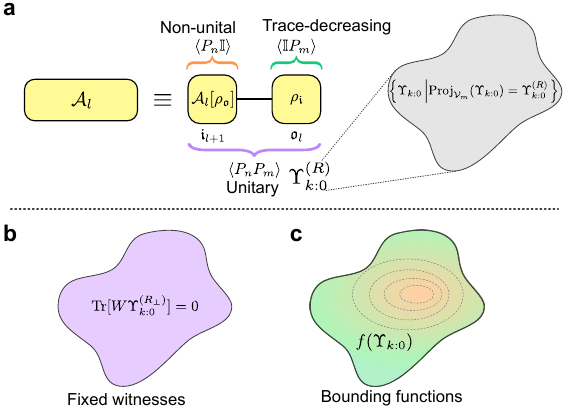}
	\caption[Illustration of various properies that can be determined from restricted process tensors ]{Property determination from restricted process tensors. \textbf{a}
		Observable interventions naturally partition into non-unital, trace-decreasing, and unitary controls. Unitary-only characterisation generates a family of process tensors consistent with this limited data. \textbf{b} Temporal entanglement may be certified by finding entanglement witnesses whose value is determined only by the observed data. Additionally, incoherent causal breaks also lie in the span of unitary operations and can be used to test non-Markovian correlation. \textbf{c} More informative functions of the process may be bounded by finding extrema across the process tensor family.
		}
	\label{fig:RPT}
\end{figure}

We will denote the space of observables taking this form by $\mathfrak{U}$, which are denoted in Fig.~\ref{fig:RPT}a. There we also list the complement of  $\mathfrak{U}$ -- i.e., non-unital and trace-decreasing control. In Sec.~\ref{sec:temp-entanglement}, we will use the form of observables belonging to $\mathfrak{U}$ to devise genuine multipartite temporal entanglement witnesses, which are depicted in Fig.~\ref{fig:RPT}b. There we will also argue that such a witness could be useful for a quantum sensor to determine whether its environment is quantum or classical. In Sec.~\ref{sec:bounding}, we use samplings of observables belonging to $\mathfrak{U}$ to extrapolate samplings from observables that lie outside of the span of $\mathfrak{U}$. That is, we extrapolate complete process tensors from samplings of a limited set of control operations. This is depicted in Fig.~\ref{fig:RPT}c.

\section{Unitaries as Temporal Entanglement Witnesses}\label{sec:temp-entanglement}

In continuing analogy with properties of states, temporal entanglement may be identified via its process tensor Choi state through any of the number of mechanisms in entanglement literature.
However, absent a unique or complete estimate for the process tensor, it is unclear whether genuinely quantum temporal correlations can be recognised. Here, we show by construction that the data generated by sequences of unitary operations surprisingly suffice in many instances to detect both bipartite and \textit{genuine multipartite entanglement} (GME) in time. We derive a restricted entanglement monotone constructed from unitary-only observables of the process. This lower bounds the generalised multipartite negativity, introduced in Ref.~\cite{PhysRevLett.106.190502}. Then, using this tool, we show that NISQ devices -- even with restricted control -- can probe multi-time physics. We conclude with a demonstration of how this fact could be significant to the field of quantum sensing, by proposing new information that a quantum sensor can detect about its environment.

\subsection{Genuine Multipartite Entanglement Witnesses from Unitary Gates}

First, we present a brief background on GME measurements following the methods introduced in Ref.~\cite{PhysRevLett.106.190502}. Consider a three-qubit state $\rho_{ABC}\in\mathcal{B}(\mathcal{H}_A\otimes\mathcal{H}_B\otimes \mathcal{H}_C)$. This state is called \emph{separable} with respect to a bipartition $A\mid BC$ if it can be written as a convex mixture of product states across that bipartition: 
\begin{equation}
	\rho_{A\mid BC}^{\text{sep}} = \sum_i p_i |\psi_i^{A}\rangle\!\langle \psi_i^{A}|\otimes |\phi_i^{BC}\rangle\!\langle \phi_i^{BC}|,
\end{equation}
and similarly across the second two bipartitions. A state which is not separable with respect to a given bipartition is said to be \emph{bipartite entangled} across that partition. If a state can be written as a convex mixture of separable states 
\begin{equation}
	\rho^{\text{bs}} = p_1\rho_{A\mid BC}^{\text{sep}} + p_2\rho_{B\mid AC}^{\text{sep}} + p_3\rho_{C\mid AB}^{\text{sep}}
\end{equation}
is said to be \emph{biseparable}. A state which is not biseparable is \emph{genuinely multipartite entangled} (GME), and is a property that is notoriously difficult to characterise except in extreme cases. 

It is well-known that separable states have a positive partial transpose (PPT). That is, let $(\cdot)^{\Gamma_A}$ be the partial transpose with respect to partition $A$. Then, for separable states the matrix $\rho_{ABC}^{\Gamma_A}$ will have only positive eigenvalues. Conversely, it is known that if $\rho_{ABC}^{\Gamma_A}$ has negative eigenvalues, then it is biparitite entangled across this partition. This is a sufficient criterion for bipartite entanglement, but is only also necessary in the case of qubit-qubit and qubit-qutrit systems.\footnote{Entangled PPT systems have a weak form of entanglement known as \emph{bound} entanglement, from which no Bell states may be distilled using local operations and classical communication~\cite{horodecki2009quantum}.} 

Ref.~\cite{PhysRevLett.106.190502} considers the characterisation of a slightly weaker form of {GME}. They define PPT mixtures to be states which can be written 
\begin{equation}
	\rho^{\text{p mix}} = p_1 \rho_{A\mid BC}^{\text{PPT}} + p_2 \rho_{B\mid AC}^{\text{PPT}} + p_3 \rho_{C\mid AB}^{\text{PPT}}. 
\end{equation}
This is slightly stronger than the biseparability criterion, and so any state which cannot be written in this form is {GME}. The advantage of this is that the set of PPT mixtures is convex, and can be fully characterised through the use of a \textit{semidefinite program} (SDP). SDPs are optimisation problems where a linear objective function over a field of matrices is minimised subject to positivity and affine constraints. These can be solved efficiently with convergence guarantees~\cite{vandenberghe1996semidefinite}.

Entanglement is typically studied through the use of entanglement witnesses. An entanglement witness $W$ is a Hermitian operator that takes a positive expectation value on all biseparable states, and a negative value on at least one entangled state. It is known that for every entangled state, one can find a witness $W$ that certifies its entanglement. However, discovery of witnesses in practice can be quite challenging due to the complicated structure of the set of biseparable states. This motivates the relaxation to finding converse states to the set of PPT mixtures. Ref~\cite{PhysRevLett.106.190502} shows that a witness $W$ of {GME} can be found for a given set of partitions by the following {SDP}:
\begin{equation}
\label{eq:ent-sdp}
	\begin{split}
	&\min \Tr[W\rho]\\
	&\text{s.t. }\Tr[W] = 1\quad \text{and} \quad \forall \ M:
	W = R_M + Q_M^{\Gamma_M};\: R_M\succcurlyeq 0,\: Q_M\succcurlyeq 0.
	\end{split}
\end{equation}
Note that GME is defined with respect to the set of partitions, and the symbol $\succcurlyeq 0$ denotes positivity of the matrices. Here, each $M$ denotes a partition of the subsystems of $\Upsilon_{k:0}$, and $\Gamma_M$ is the partial transpose with respect to that partition. The set of all available partitions will depend on the target of the entanglement witness -- i.e. an $n$-partite entanglement witness will cover $2^{n-1}-1$ unique partitions\footnote{A set of size $n$ has a power set of size $2^n$, but only half of these need to be checked due to the symmetry of the partial transpose, and we can exclude the null set.}. If $W$ is found such that $\text{Tr}[W\Upsilon_{k:0}] < 0$ then $\Upsilon_{k:0}$ is genuinely multipartite entangled across the defined partitions. 
$W$ and each $R_M$ are the free parameters of the optimisation, equivalently imposing that $(W - R_M)^{\Gamma_M}\succcurlyeq 0$. The role of $R_M$ and $Q_M$, therefore, is to maintain PPT of the witness. If the minimum of this optimisation is negative, then $\rho$ is not a PPT mixture, and hence {GME}. 

The above results straightforwardly generalise to the temporal case because the process tensor is just a many-body state. Operationally, temporal entanglement means that the environment is genuinely a quantum memory, and when it is GME the process has the ability to transform bipartite entanglement into multipartite entanglement~\cite{milz21}. This is a common feature in complex quantum systems, namely turning local coherence into global ones. For instance, quantum chaos leads to highly entangled spatiotemporal states~\cite{PRXQuantum.5.010314}.

Naturally, in the context of multi-time processes, we identify $\rho$ with a process tensor $\Upsilon_{k:0}$.
The objective function $\text{Tr}[W\Upsilon_{k:0}]$ constitutes a prediction of the expectation of $W$ with respect to $\Upsilon_{k:0}$ if it were to be measured in the lab. If the instruments $\{\mathcal{B}_{j}^{\mu_j}\}$ used to reconstruct $\Upsilon_{k:0}$ are not IC, then we can only make predictions with confidence with respect to the support of $\{\mathcal{B}_{j}^{\mu_j}\}$. That is, we require $W\in\mathfrak{U}$. Since this is an affine constraint, it can be added to the SDP~\eqref{eq:ent-sdp} to find an efficient solution. A straightforward way to do this is to construct a basis for $\mathfrak{U}^{\perp}$ and demand that $W$ be individually orthogonal to each element of that basis. In our context, we pick this basis to be all Pauli strings not of the form of Eq.~\eqref{eq:restr-obs}.
The following modified SDP hence produces a restricted entanglement witness for processes:
\begin{equation}
	\label{eq:SDP-witness}
	\begin{split}
		&\min\text{Tr}[W\Upsilon_{k:0}] \\ & \text{s.t.} \ 
  \text{Tr}[W] = 1, \ \text{Tr}[W\mathsf{P}]=0 \ \forall \ \mathsf{P}\in \mathfrak{U}^{\perp}, \\ & \text{and} \ \forall \ M : W = R_M + Q_M^{\Gamma_M}; R_M\succcurlyeq 0, \ Q_M\succcurlyeq 0.
	\end{split}
\end{equation}

We introduce one final modification. In Ref.~\cite{PhysRevLett.106.190502} it is shown that if the condition $\text{Tr}[W] = 1$ is replaced with $R_M, Q_M \preccurlyeq \mathbb{I}$ then the resulting minimisation defines an entanglement monotone, which we employ. This quantity has several desirable properties, including take a zero value for biseparable states and being non-increasing under full LOCC. In the bipartite case, it reduces to the negativity. Here, since we restrict the span of witnesses, we can only lower bound the monotone.
Witnessing temporal entanglement certifies that the non-Markovian memory is genuinely quantum~\cite{Giarmatzi2021,milz21}-- i.e., it's a tool that gives us a peek into the nature of the environment.

When a restricted process tensor $\Upsilon_{k:0}$ is estimated from limited experimental data, a non-unique Choi state is constructed obeying conditions of complete positivity and causality. But so long as $W\in\mathfrak{U}$ is imposed, the witness is restricted to be an observable on $\Upsilon_{k:0}^{(R)}$ which \textit{is} uniquely fixed. Thus, we may solve the {SDP} given in Eq.~\eqref{eq:SDP-witness} to detect both bipartite and multipartite entanglement in time, even with extremely simple circuits such as sequences of unitary operations. In practice, we use the Python convex optimisation package \texttt{PICOS} with the \texttt{MOSEK} solver~\cite{PICOS,mosek}. Although this witness is not as powerful as a fully generic observable on $\Upsilon_{k:0}$, we find it sufficient to detect temporal entanglement in practice. This result extends the boundaries of what can be determined about the dynamics of an open quantum system, and should in particular find utility in the study of naturally occurring quantum stochastic processes. \par

\subsection{Temporal Entanglement Witness Demonstration on NISQ Devices}

\begin{figure}
	\centering
	\includegraphics[width=0.95\linewidth]{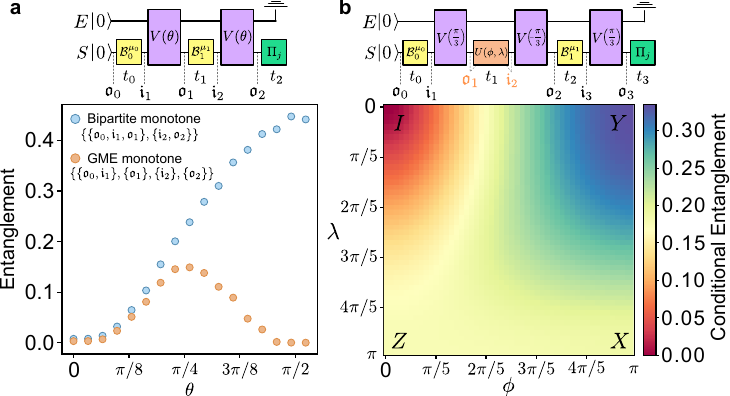}
	\caption{Implementation of unitary-restricted temporal entanglement witnesses. \textbf{a} Results of a circuit designed to manufacture different temporal entanglement structures implemented on \emph{ibm\_cairo}. The quantity plotted is the absolute value of the solved SDP in Eq.~\eqref{eq:SDP-witness} with respect to the process tensor Choi state. \textbf{b} We construct a three-step process tensor on \emph{ibm\_auckland} and show how deterministically changing the intermediate unitary gives control over the resulting temporal quantum correlations. The quantity plotted is again the solution to Eq.~\eqref{eq:SDP-witness}, but with respect to the conditional process tensor Choi state, defined in Eq.~\eqref{eq:conditional-pt}.}
	\label{fig:ent-witnesses}
\end{figure}

To demonstrate that our entanglement characterisation is relevant for near-term experiments, we consider two settings on IBM Quantum devices \emph{ibm\_cairo} and \emph{ibm\_auckland}. 
With one qubit as the simulated environment and a two-step (three-time) process tensor, we tune the $SE$ interaction between two qubits to $V(\theta):= \exp\left(-i\frac{\theta}{2}\sum_iP_i\otimes P_i\right)$, and determine both the detectable bipartite and genuinely multipartite entanglement as a function of $\theta\in [0,\pi/2]$. The start, middle, and end of these values respectively produce an identity, $\sqrt{\text{SWAP}}$, and SWAP gate. These interactions are designed to produce different temporal entanglement structures. Repeated $\sqrt{\text{SWAP}}$ gates in principle produce {GME} across legs $\{\mathfrak{i}_1, \mathfrak{o}_1, \mathfrak{i}_2, \mathfrak{o}_2\}$~\cite{Milz2021PRXQ}, see circuits in Figure~\ref{fig:ent-witnesses}. Meanwhile SWAPs ought to produce maximal bipartite entanglement between $\hat{\mathcal{E}}_{1:0}$ and $\hat{\mathcal{E}}_{2:1}$, since $\mathfrak{i}_1$ to $\mathfrak{o}_2$ is an effective identity channel -- but $\mathfrak{o}_1$ and $\mathfrak{i}_2$ are uncorrelated, hence no {GME}.

We realise and quantify these entanglement structures using our entanglement bound which relies only on unitary sequences. We perform MLE-PTT to reconstruct the restricted process tensor for the above scenario. This requires 300 different measurement settings: combinations of 10 unitaries at $t_0$, 10 unitaries at $t_1$, and 3 measurement bases at $t_2$. Note that this assumes that the single-qubit unitary gates of the device have negligible gate-dependent noise. Given that these total error rates are $\mathcal{O}(10^{-4})$, we expect the gate-dependent components to be orders of magnitude smaller than the signal we are measuring. Nevertheless, the effects of gate imperfection can be accounted for~\cite{Li_2024, white2023unifying}.
From the resulting estimate $\Upsilon_{2:0}^{(R)}$ we then solve the SDP~\eqref{eq:SDP-witness} to produce a lower bound on the entanglement. Although we only have access to the restricted process tensor, solving this problem tells us about the entanglement of the full process. The results of this are shown in Figure~\ref{fig:ent-witnesses}a, where we can see how the temporal entanglement is tuned from a trivial to a multipartite, and then a bipartite structure as a function of the interaction strength $\theta$. This demonstrates the ability to robustly detect entangled multi-time processes on real hardware even with limited control.

In addition to varying the interaction, we also show that for a fixed process, one can `inject' complexity into the circuit~\cite{aloisio-complexity} via local control. We consider a three-step process across times $\{t_0,t_1,t_2,t_3\}$. If an operation is applied at $t_1$, then a temporal structure is induced across $\{t_0,t_2,t_3\}$. This structure will depend on the experimenter's choice. Our setup, then, is to consider once more a single qubit system with a single qubit environment, this time on \emph{ibm\_auckland}. The three steps consist fixed interactions $V(\pi/3)$, $V(-\pi/3)$ and $V(\pi/3)$ between the two qubits. This setup is depicted in the circuit of Figure~\ref{fig:ent-witnesses}b. As above, we reconstruct the three-step process tensor $\Upsilon_{3:0}^{(R)}$ using MLE-PTT. We can use this to determine the structure on $\{t_0,t_2,t_3\}$ on the application of some parametrised unitary $U(\phi,\lambda)$ at $t_1$. Then, for varied parameter values, compute the certifiable bipartite entanglement between $\{\mathfrak{o}_0,\mathfrak{i}_1,\mathfrak{o}_2\}$ and $\{\mathfrak{i}_3,\mathfrak{o}_3\}$. Thus, process correlations may be greatly manipulated even at the system level.
Applying a fixed operation on the system physically transforms one process into a new one. This new process can be taken to be the old process, where one step is projected onto the fixed operation. In our example of controllable entanglement, Figure~\ref{fig:ent-witnesses}b, we first estimate a three-step process tensor $\Upsilon_{3:0}^{(R)}$. A conditional two-step marginal is then found for parametrised unitary
\begin{equation}
	U(\phi,\lambda) = \begin{pmatrix}
		\cos(\phi/2) & -\text{e}^{i\lambda}\sin(\phi/2) \\
		\text{e}^{i\lambda}\sin(\theta/2) & \cos(\phi/2)
	\end{pmatrix}.
\end{equation}
Let $\hat{\mathcal{U}}(\phi,\lambda)$ be the Choi state of $U(\phi,\lambda)$, then
\begin{equation}
\label{eq:conditional-pt}
	\Upsilon_{2:0|U(\phi,\lambda)} = \text{Tr}_{\mathfrak{i}_2,\mathfrak{o}_1}\left[(\mathbb{I}_{\mathfrak{o}_3,\mathfrak{i}_3, \mathfrak{o}_2}\otimes \hat{\mathcal{U}}(\phi,\lambda)^\text{T}\otimes \mathbb{I}_{\mathfrak{i}_1,\mathfrak{o}_0})\cdot\Upsilon_{3:0}\right],
\end{equation}
which is the two-step conditional marginal process where $U(\phi,\lambda)$ is applied in position 1 of the circuit. For $\phi\in[0,\pi]$ and $\lambda\in[0,\pi]$ we compute the unitary entanglement lower bound Eq.~\eqref{eq:SDP-witness} between legs $\{\mathfrak{o}_0,\mathfrak{i}_1\}$ and $\{\mathfrak{o_2},\mathfrak{i}_3,\mathfrak{o}_3\}$. Where $U$ is taken to be the identity, there is naturally no entanglement present in the remaining dynamics. However, as the unitary is varied, the effective interaction from $t_0$ to $t_2$ becomes complex, generating temporal entanglement.

What this is intended to emphasise is that an experimenter can tune the complexity dynamics based with their choice of control operations. Above, one choice leads to multi-time entanglement, while another leads to two-time entanglement, which can be related to the complexity of quantum dynamics~\cite{aloisio-complexity}, including for quantum master equations~\cite{pollock-tomographic-equations}. We explore in the next section how temporal entanglement emerges in sensing, and how quantum sensors -- which have natural control constraints -- can be used to learn more about the environment than was previously realised.

\subsection{Quantum Sensing of Quantum Correlated Processes }

The ideas set out in the previous subsection have a direct application for probing correlated environments. Here, we will explore this for quantum sensors, but the same ideas apply to qubits belonging to a quantum computer. Quantum sensors are controllable quantum systems exploited to perform a measurement of some external quantity, referred to as the \emph{signal}~\cite{RevModPhys.89.035002}. The typical sensing protocol is (a) to initialise the quantum sensor into a desired sensing state $\ket{\psi}$ (usually a superposition), (b) allow the sensor to evolve under the Hamiltonian $H_{SE}$ for time $\tau$, (c) apply a control operation $U_c$ to the state $\ket{\psi(\tau)}$, and (d) perform a projective readout. Steps (b) and (c) can be repeated for a multi-control sequence; the evolution time $\tau$ may be varied; and the entire cycle will be repeated many times to collect statistics. This is the typical set of controls assumed of a sensor: reliable initialisation, precise timing, robust unitary control, and a terminating measurement. For example, an external environment may cause some accumulated phase on the system which can be observed as a difference in $\ket{0}$ and $\ket{1}$ populations after a basis change. 

A typical use case consists of environment spectral estimation, where \emph{dynamical decoupling} (DD) (otherwise known as multi-pulse sensing) sequences take the position of (b) and (c) in the above protocol~\cite{RevModPhys.89.035002}. In the context of quantum information, DD sequences are typically known for prolonging the coherence time of a system by decoupling it from its decoherence modes~\cite{gullion1990new, Viola-DD}. However, DD may additionally be understood as in terms of a weighting function that modulates different frequencies and amplitudes, also known as a \emph{filter function}. In periodic DD, for example, these weighting functions behave like narrow-band filters around a centre frequency and its harmonics. For a large number of pulses, then, these transmission frequencies are amplified and other frequencies are suppressed. Periodic DD in particular can have its pass-band filter tuned by varying the evolution time $\tau$. In effect, this approach allows one to `lock' onto a part of the environment and amplify its signal. This is commonly used in, for example, the sensing of nearby nuclear spins using nitrogen-vacancy (NV) defects~\cite{staudacher2013nuclear}. We show that this feature, in conjunction with restricted PTT, allows one to determine not only this environment signal, but also its classical or quantum nature.

\begin{figure}[t!]
	\centering
	\includegraphics[width=\linewidth]{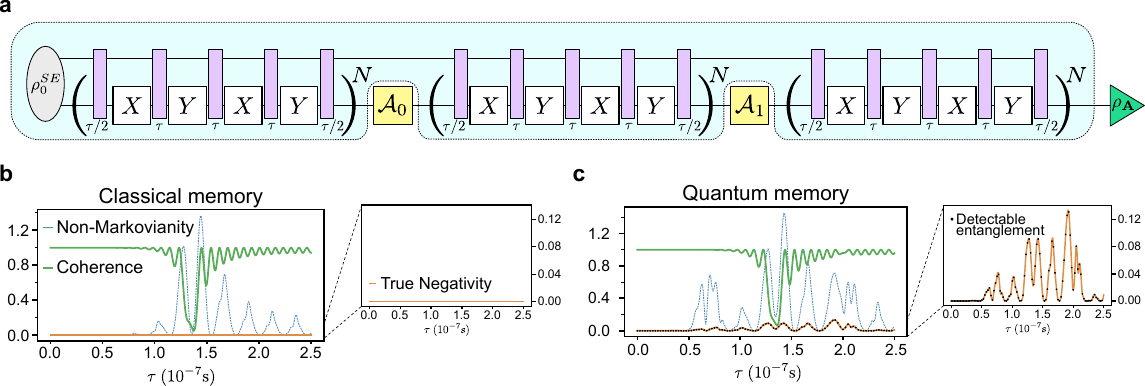}
	\caption{Implementation of unitary-restricted temporal entanglement witnesses in the context of quantum sensors. 
	\textbf{a} A circuit diagram illustrating a process tensor built around a DD sequence. When on resonance, the DD sequence amplifies interactions with the environment. The process tensor can then be used to detect and characterise the temporal correlations generated. For comparison, we also plot total non-Markovianity which is the quantum mutual information of the Choi state.
	\textbf{b} 
    A quantum environment with locally diagonalisable interactions are equivalent to a purely classical memory. Hence, these dynamics will generate non-Markovian correlations, but entanglement in the process tensor of a simulated quantum sensor remains zero. \textbf{c} When the multi-time correlations are genuinely quantum, we show that this may be detectable by the sensor with unitary-constrained control. The green curve represents the conventional sensing information, and the black dots represent the information obtainable through our restricted unitary witness approach.}
	\label{fig:sensor-entanglement}
\end{figure}

In App.~\ref{app:sensor}, we show that the process tensor must be temporally separable when the environment is classical. A classical environment is nothing more than a time-fluctuating field, which always leads to a process tensor that is a convex mixture of unitary processes. In contrast, when the environment is a quantum system itself, the corresponding process tensor may have temporal entanglement. 
Quantumness of the environment, however, is a necessary condition and not a sufficient one to guarantee temporal entanglement. In App.~\ref{app:sensor} we show this by with the additional constraint that the interaction cannot be diagonalisable via local operations on $S$ and $E$. Our definition of classical in this instance is hence a set of temporal correlations that could be classically emulated, but these may stem from a quantum system.
Importantly, these two scenarios are not distinguishable via standard sensing methods, which works just as well in either case.

Here, we will use the filter functions induced by DD sequences. By varying the evolution time $\tau$, we can sweep through different frequencies, and the effective filter function will act as a modulator of the signal.
When the frequency of the environment matches the transmission frequency of the sequence, the interaction is amplified and system will decohere. The coherence is determined via a simple population measure. First, initialise the system into the state $\ket{i+} = \tfrac{1}{\sqrt{2}}
(\ket{0} + i\ket{1})$, then apply the DD sequence with wait time $\tau$, finally, apply an $R_x(-\pi/2)$ gate and determine the $\langle Z \rangle$ expectation value. We propose the reconstruction of a restricted process tensor around the {DD} sequence. The use of our unitary temporal entanglement witnesses can then determine whether the interaction is genuinely quantum or not.
The described process tensor is depicted in Figure~\ref{fig:sensor-entanglement}a, where two steps are interleaved between an $XY4$ sequence repeated $N$ times (in the numerics, $N=3$). 
By constructing unitary-restricted process tensors within the filter, we show that quantumness of the interaction may be certified by measuring entanglement in the multi-time correlations. This shows how one in principle may operationally distinguish between different classes of Hamiltonian dynamics, even for open quantum systems. Our procedure is entirely consistent with the capabilities of quantum sensors, and adds a different capacity to learn about the system-environment interaction through structured sequences on the system.

We demonstrate this idea numerically. The simulation takes a simple spin-spin Hamiltonian $H = H_S\otimes H_E$ in the rotating frame of the sensor. This has the form:
\begin{equation}
   H = I^{(S)}\otimes\omega_LZ^{(E)} + H_I,
\end{equation}
where $\omega_L$ is the Larmor frequency of the environment spin, which is modelled as a nuclear spin with frequency $\omega_L/2\pi = 1.85\times 10^6$ Hz, 
and $H_I$ is the interaction Hamiltonian modelling some possible hyperfine coupling.
For the examples of classical and quantum interactions, we choose interactions:
\begin{equation}
    H_I^{(\text{Classical})} =  g\cdot Z\otimes(X+Z)
    \qquad \text{versus} \qquad
    H_I^{(\text{Quantum})} = g\cdot(Z\otimes Z + Z\otimes X + X\otimes X),
\end{equation}
where $g=1.1\times10^5$ Hz. 
The interaction $H_I^{(\text{Classical})}$ generates no entanglement in the process tensor because it can be locally diagonalised (see App.~\ref{app:sensor}). Thus, after tracing over the environment, the interaction is locally equivalent to a stochastic mixture of $Z-$rotations, which is non-Markovian but separable. It could hence be replaced with a classical source of correlated randomness, such as a fluctuating magnetic field. In contrast, $H_I^{(\text{Quantum})}$ is a genuinely quantum memory whose correlations could not be spoofed by a classical agent.
Here, we can distinguish these two scenarios through purely system-level quantities. 

We first simulate the coherence curves of the different interactions, as determined by an $XY4$ frequency sweep. Roughly speaking, the environment frequency determines the location of the dip (the resonance for PDD is at $\tau_0=\pi/\omega_L$), and the coupling strength to first order determines its amplitude. Each of these effects (in addition to some harmonic frequencies) are shown in the coherence curves of Figure~\ref{fig:sensor-entanglement}b and c. Since both classical and quantum baths are capable of decohering a system, we do not obtain any additional information from these curves. Instead, we reconstruct our process tensor across the DD sequence. For each $\tau$, we can compute our restricted entanglement witness. In the case of $H_I^{(\text{Classical})}$, this witness is always zero. Meanwhile, $H_I^{(\text{Quantum})}$ generates interactions that do not commute between different times. It hence verifiably generates quantum correlations in time, which could only be created by a genuinely quantum environment. These results are shown in Figures~\ref{fig:sensor-entanglement}b and c. For completeness, we also plot the total non-Markovianity as measured by the quantum mutual information (see this next section). This quantity is not measurable under the present conditions, but indicates that temporal correlations are indeed mediated by the bath in both instances.

\section{Bounding Process Properties}
\label{sec:bounding}

The previous section was concerned with witnesses of \textit{quantum} non-Markovianity which could be found entirely through the span of unitary operations. In general, this is insufficient to describe both classical correlations and more sophisticated measures of non-Markovianity with operational interpretations~\cite{Milz2021PRXQ}. Moreover, we are only able to bound temporal entanglement from below~\cite{White-NM-2020}, and thus the complementary upper bounds, with some tightness guarantees, are desirable. Measures of non-Markovianity typically require access to the full Choi state, including its spectrum. This cannot be uniquely reconstructed in the tomographically incomplete regime. That is, given restricted tomography data, it is not possible to directly estimate non-Markovian correlations.

Here, we develop a method that allows for both the lower and upper bounding of properties of a non-uniquely reconstructed process tensor. From this, we show how it is possible to obtain various detailed properties of quantum processes, including purity, fidelity, and generalised quantum mutual information. This permits the study of multi-time physics on both quantum computers either without or with unreliable mid-circuit measurement capabilities and, importantly, quantum sensors which may be placed in highly complex environments. 

We circumvent the limitation of unitary control by constructing regions of plausibility and bounding process properties among these. Measuring the outcomes to sequences of unitaries naturally generates a `family' of process tensors which are consistent with the incomplete data. This idea is sketched in Figure~\ref{fig:RPT}c. We develop an algorithm that searches this space of plausible process tensors to find bounds over objective functions of the user's choosing. Thus, even without a complete estimate of a multi-time process, one may answer the next best question: what range of dynamics is consistent with a limited set of observations?

Note that the methods developed here are for accessing multi-time non-Markovian correlations. Two-time non-Markovian measures, e.g. those presented in Refs.~\cite{PhysRevLett.103.210401, PhysRevLett.105.050403}, do not require mid-circuit measurements. However, such measures of non-Markovianity are not complete and can be seen as limiting cases of the process tensor~\cite{Pollock2018}. Importantly, the process tensor quantifies correlations that can be shown to be relevant for quantum control~\cite{White-MLPT, berk2021extracting}. Thus, the results below allow for estimating multi-time non-Markovian memory, even when limited to a restricted control set.

\subsection{Searching Families of Process Tensors}
Suppose an experimenter wished to construct a maximum-likelihood model of their data which explained why sequences of operations $\mathbf{B}_{k-1:0}^{\vec \mu}$ led to observed frequencies $m_{\vec{\mu}}$. The process tensor model $\Upsilon_{k:0}$ is a positive semi-definite operator that obeys a set of causality conditions, which, as we have seen, is enforced by $\Upsilon_{k:0}\in \mathcal{S}_n^+\cap \mathcal{V}_{\rm c}$, where $\mathcal{S}_n^+$ is the cone of PSD matrices, and $\mathcal{V}_{\rm c}$ is the hyperplane of causal constraints. When the set of sequences is informationally complete, the maximum-likelihood estimate is a unique point in this set. However if the sequence of operations is solely in the span of unitaries, then the maximum-likelihood estimate is non-unique: there is a hyperplane of process tensors through which the model can be varied without changing the likelihood function. Since we have no extra data through which to add credence to any of these models, the best we can do is explore the space to evaluate models which are plausible with the data. Specifically, we look at exploring the values of different objective functions over this space. Global maxima and minima will respectively correspond to upper and lower bounds for the objective function as fixed by the restricted set of data. Since the set is convex, then if the function is also convex these bounds will be efficiently computable. Otherwise it is a harder (but, nevertheless, tractable\footnote{Although non-convex optimisation is NP-hard in general, the size of the problem becomes bottle-necked by the quantum computational resources long before the classical demands.}) optimisation problem. 

We introduce our bounding algorithm by noting that the set of observed measurement outcomes forms a hyperplane constraint. Consider a process tensor with decomposition $\Upsilon_{k:0} = \Upsilon_{k:0}^{(R)} + \Upsilon_{k:0}^{(R_\perp)}$. Here, 
$\Upsilon_{k:0}^{(R)}$ is the a restricted process tensor and $\Upsilon_{k:0}^{(R_\perp)}$ is an element from the orthogonal complement of $\Upsilon_{k:0}^{(R)}$ such that the total process tensor is physical, i.e., satisfying Eq.~\eqref{eq:processtensor}. In general, given a restricted process tensor $\Upsilon_{k:0}^{(R)}$ -- which will be fixed by the MLE step -- there is a family of complement elements $\{ \Upsilon_{k:0}^{(R_\perp)} \}$ that will yield a family of physical full process tensors $\{\Upsilon_{k:0}\}$. Our goal will be to characterise and explore the properties of this family.

Our starting point is the limited control set with basis size $N_r < d^4$
\begin{equation}
\left\{\mathbf{B}_{k-1:0}^{\vec{\mu}}\right\} = \left\{\bigotimes_{i=0}^{k-1}\mathcal{B}_i^{\mu_j}\right\}_{\vec{\mu}=(1,1,\cdots,1)} ^{(N_r,N_r,\cdots,N_r)},
\end{equation}
and the corresponding experimental data, $m_{i,\vec{\mu}}$. From this, an MLE estimate $\Upsilon_{k:0}^{(\text{MLE})}$ is found, from which we obtain an estimate of the true probabilities:
\begin{equation}\label{eq:ptdata}
    \text{Tr}\left[\Upsilon_{k:0}^{(\text{MLE})} \left(\Pi_i\otimes \mathbf{B}_{k-1:0}^{\vec{\mu}\text{T}}\right)\right] = p_{i,\vec{\mu}}.
\end{equation}
Eq.~\eqref{eq:ptdata} defines an affine space through the set of linear equations satisfied by $\Upsilon_{k:0}^{(\text{MLE})}$. We denote this affine space as $\mathcal{V}_{\rm m}$ and append it to the set of causality conditions, i.e., the set of Pauli operator strings $\mathsf{P}_i \in \mathfrak{C}$ whose expectation values must vanish as per Eq.~\eqref{eq:causalityPauli}. Together this describes the affine space $\mathcal{V} := \mathcal{V}_{\rm c}\oplus \mathcal{V}_{\rm m}$, which can be more explicitly written as the set of $\Upsilon_{k:0}$ such that
\begin{gather}
	\label{affine-constraint-bound}
		\begin{pmatrix} \langle\!\langle \mathsf{P}_0| \\
		\vdots \\
		\langle\!\langle \mathsf{P}_N|\\
		\langle\!\langle\mathbb{I}|\\
		\langle\!\langle{\Pi_{i_0}\otimes\mathbf{B}_{k-1:0}^{\vec{\mu}_0}}|\\
		\langle\!\langle{\Pi_{i_t0}\otimes\mathbf{B}_{k-1:0}^{\vec{\mu}_1}}|\\
		\vdots \\ 
		\langle\!\langle{\Pi_{i_L}\otimes\mathbf{B}_{k-1:0}^{\vec{\mu}_N}}|\\
		\end{pmatrix} \cdot |\Upsilon_{k:0}\rangle\!\rangle = \begin{pmatrix} 0\\\vdots\\0\\1\\p_{i_0,\vec{\mu}_0}\\p_{i_0,\vec{\mu}_1}\\\vdots\\p_{i_L,\vec{\mu}_N}\end{pmatrix}.
\end{gather}

Recall that the MLE problem is to find $\Upsilon_{k:0}\in\mathcal{S}_n^+\cap\mathcal{V}_{\rm c}$ such that the likelihood function $f(\Upsilon_{k:0})$ in Equation~\eqref{eq:mle} is minimised. Ref.~\cite{White-MLPT} developed a projected gradient descent routine based on the results of~\cite{QPT-projection} to solve this problem. This is known as the \texttt{pgdb} algorithm. We now extend this in two ways: first, we expand the affine space from $\mathcal{V}_{\rm c}$ to $\mathcal{V}_{\rm c}\oplus\mathcal{V}_{\rm m}$; and second, we consider a range of different information-theoretic quantities, labelled generically as $\mathcal{I}$ as our objective to find maxima and minima. This opens up the exploration of process tensor families consistent with the limited observations. Using this technique we find minimum and maximum bounds for the values that any differentiable function can take for all process tensors consistent with observed data. The gradient of the function may be supplied analytically, computed through automatic differentiation libraries, or calculated numerically through the method of finite differences. 

Full details and benchmarking of the \texttt{pgdb} algorithm for QPT can be found in Ref.~\cite{QPT-projection}. For completeness, Algorithm~\ref{alg:pgdb} details the \texttt{pgdb} pseudocode. In short, the process tensor is initialised as its MLE estimate $\Upsilon_{k:0}^{\rm (MLE)}$. Each step of the optimisation consists of (a) a gradient step in the direction that minimises $\mathcal{I}$, (b) a projection step, such that the minimisation is only in the direction where $\Upsilon_{k:0} \in\mathcal{S}_{n}^+\cap\mathcal{V}$, and (c) a backtracking step to ensure that the step is not too large. Algorithmic performance of the projection subroutine is contingent on the condition number of the constraint matrix. For this reason, it is important to cast the problem not necessarily in terms of the Choi states of the experimental operations, but ideally in terms of a mutually unbiased basis that spans the same space. In our case, with unitary-only control, these were the weight-2 Pauli operators on $\mathscr{B}(\mathcal{H}_{\text{out}})\otimes \mathscr{B}(\mathcal{H}_{\text{in}})$.

\begin{algorithm}[t]
\caption{Exploring Process Tensor Families}
\label{alg:pgdb}
\begin{algorithmic}[1]
\State $j=0,n=d_S^{2k+1}$
\State \text{Initial estimate}:
$\Upsilon_{k:0}^{(0)} = \Upsilon_{k:0}^{\text{(MLE)}}$ \Comment{Initialises $\Upsilon_{k:0}$ with its non-unique restricted estimate.}
\State \text{Set metaparameters: }$\alpha=2n^2/3,\gamma=0.3,\mu=1\times 10^{-3}$\Comment{Responsibe for gradient and backtrack step sizes.}
\While{$\mathcal{I}(\Upsilon_{k:0}^{(j)}) - \mathcal{I}(\Upsilon_{k:0}^{(j+1)}) > 1\times 10^{-6}$} \Comment{Convergence criterion.}
\State $D^{(j)} =\text{Proj}_{S_n^+\cap \mathcal{V}}\left(\Upsilon_{k:0}^{(j)} - \mu \nabla \mathcal{I}(\Upsilon_{k:0}^{(j)})\right) - \Upsilon_{k:0}^{(j)}$\Comment{Projected gradient step, maintains positivity, causality, and consistency with the restricted data.}
\State $\beta = 1$
\While{$\mathcal{I}(\Upsilon_{k:0}^{(j)}) + \beta D^{(k)}) > \mathcal{I}(\Upsilon_{k:0}^{(j)}) + \gamma \beta \left\langle D^{(j)}, \nabla \mathcal{I}(\Upsilon_{k:0}^{(j)})\right\rangle$} \Comment{Steps in appropriate gradient direction to reduce cost function.}
\State $\beta = 0.5\beta$
\EndWhile
\State $\Upsilon_{k:0}^{(j+1)} = \Upsilon_{k:0}^{(j)} + \beta D^{(j)}$
\State $j = j+1$
\EndWhile
\State\Return $\Upsilon_{k:0}^{(\text{est})} = \Upsilon_{k:0}^{(j+1)}$
\end{algorithmic}
\end{algorithm}

Full details of the projection subroutine
\begin{equation}
\label{proj-def}
    \text{Proj}_{S_n^+\cap\mathcal{V}}(\Upsilon) = \argmin_{\Upsilon' \in S_n^+\cap\mathcal{V}}||\Upsilon'-\Upsilon||_2^2
\end{equation}
can be found in Refs.~\cite{White-MLPT,conic-projection,conic-handbook}. For completeness, we include a brief summary here of this important subroutine. We emphasise that this is only to perform the projection, and is not related to the other steps of \texttt{pgdb}. Let $\kket{\Upsilon_{k:0}} \equiv \kket{\Upsilon}$, let $\vec{\kappa}\in\mathbb{R}^m$ for $m$ affine constraints, and let the terms in Eq.~\eqref{affine-constraint-bound} be denoted equivalently $A\cdot \kket{\Upsilon} = b$. Diagonalising $\Upsilon$ gives $\Upsilon = UDU^\dagger$ where $D = \text{diag}(\lambda_1, \lambda_2,\cdots,\lambda_n)$ is real. Then the projection onto $\mathcal{S}_n^+$ is:
\begin{gather}
    \label{PSD-proj}
    \text{Proj}_{\mathcal{S}_n^+}(\Upsilon) = U \text{diag}(\lambda^+_0,\cdots,\lambda^+_n)U^\dagger
\end{gather}
with $\lambda^+_j:=\max\{\lambda_j,0\}$. Now, defining $|\Upsilon'(\vec{\kappa})\rangle\!\rangle := \text{Proj}_{\mathcal{S}_n^+}(\kket{\Upsilon} + A^\dagger \vec{\kappa})$ (note that projection of a vector here carries implied matrix reshape, positive projection, followed by vector reshape), it can be shown that the unconstrained minimisation of the function
\begin{equation}
    \theta(\vec{\kappa}) = -\left\|\kket{\Upsilon'(\vec{\kappa})}\right\|^2 + b^\dagger \vec{\kappa},
\end{equation}
whose gradient is
\begin{equation}
    \nabla\theta(\vec{\kappa}) = -A\kket{\Upsilon'(\vec{\kappa})} + b,
\end{equation}
is the unique solution to Eq.~\eqref{proj-def}. For our purposes then, we employ the L-BFGS algorithm to minimise $\theta(\vec{\kappa})$, and thus project some $\Upsilon$ onto $\mathcal{S}_n^+\cap\mathcal{V}$. This completes the description of an algorithm to bound arbitrary properties of processes from incomplete data.

\subsection{Diagnosing Non-Markovian Noise on NISQ Devices}

\begin{figure*}[ht]
    \centering
    \includegraphics[width=0.5\linewidth]{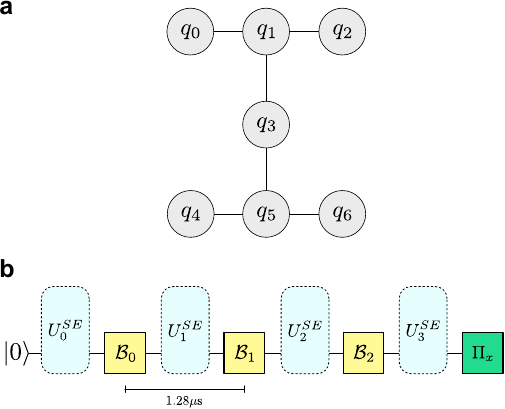} 
    
    \vspace{0.5cm}

    \resizebox{\linewidth}{!}{ 
        \begin{tabular}{@{}lcccc@{}}
            \toprule
            Setup & QMI (min, max) & Negativity & Purity & Fidelity \\ 
            \midrule
            \#1. System Alone & (0.298, 0.304) & (0.0179, 0.0181) & (0.8900, 0.8904) & (0.9423, 0.9427) \\
            \#2. $\ket{+}$ Nearest Neighbours (NN) & (0.363, 0.369) & (0.0255, 0.0259) & (0.7476, 0.7485) & (0.7239, 0.7244) \\
            \#3. Periodic CNOTs on NNs in $\ket{+}$ & (0.348, 0.358) & (0.0240, 0.0246) & (0.7796, 0.7811) & (0.7264, 0.7272) \\
            \#4. $\ket{0}$ NNs with DD on System & (0.358, 0.373) & (0.0205, 0.0210) & (0.8592, 0.8612) & (0.8034, 0.8045) \\
            \#5. $\ket{+}$ Long-range Neighbours & (0.329, 0.339) & (0.0209, 0.0214) & (0.8594, 0.8608) & (0.9252, 0.9260) \\
            \#6. $\ket{+}$ NN, delay, $\ket{+}$ next-to-NN & (0.322, 0.329) & (0.0209, 0.0213) & (0.8534, 0.8549) & (0.9077, 0.9083) \\
            \bottomrule
        \end{tabular}%
    }
    
    \caption{\textbf{a} Qubit map of \emph{ibmq\_casablanca}, $q_5$ was the system qubit in all experiments. \textbf{b} Structure of the process tensor circuits for this device: a three-step process tensor constitutes three basis operations, followed by measurement. For all six experiments, we fixed the wait time at $1.28\:\mu$s and varied the background. \underline{Table}: Property bounds on process tensors with different background dynamics. Presented figures are properties of three-step processes. {QMI} measures the non-Markovianity of the process, with conditional {QMI} the minimal and maximal past-future dependence for first and last steps. Negativity measures unbound temporal bipartite entanglement. Purity indicates purity of the ensemble of open system trajectories, while fidelity captures the level of noise.}
	\label{tab:memory-bounds}
\end{figure*}

We apply this framework in the interest of mapping out the quantum and classical temporal correlations present in naturally occurring noise where it was not possible to perform complete tomography. We constructed six three-step process tensors in different dynamical setups for a system qubit on the superconducting device \emph{ibmq\_casablanca}. See Figure~\ref{tab:memory-bounds} for a schematic of the device and circuits. These setups were designed to examine base non-Markovianity, as well as crosstalk influence. Specifically, the scenarios are as follows. Setup \#1 features only control operations on the single (system) qubit. Setup \#2 places the two geometric neighbours of the system qubit into the $\ket{+}$ state, where the natural $ZZ$ crosstalk of the device may play a role. This amounts to a more complex environment than setup \#1. Setup \#3 places these qubits once more into the $\ket{+}$ state, but subjects them and their neighbours to a repeated CNOT gate, where active crosstalk here may play a role. Setup \#4 has the system qubit subject to dynamical decoupling (DD) sequences between steps, to determine whether this plays a role in the noise on the system qubit. Applying DD sequences can, for instance, change the effective axis of the noise if it is not fully removed, which might change the memory structure. Setup \#5 places the next-to-nearest geometric neighbours into the $\ket{+}$ state, in the event that crosstalk persists more non-locally (the distinction with setup \#2 is the $\ket{+}$ initialisation of qubits one neighbour further). Lastly, setup \#6 looks at the interplay between combining setups \#2 and \#5, where the effective bath is larger. Across these series of experiments, our intention was to probe the interplay between naturally present memory effects at the system level and those which are driven by potential crosstalk interactions. In all cases, the two-qubit gates are applied to the neighbouring qubits, and not the system.
More information about the device and the specifics of the experiment can be found in Appendix~\ref{app:exp}.
A ten-unitary basis was used (minimal complete for single-qubit unitaries, see Sec.~\ref{ssec:partitioning-control}), and each process tensor obtained with maximum-likelihood {PTT}. Fixing these observations in the model, we then searched the family of consistent process tensors for a variety of diagnostic measures.

Specifically, we consider four information-theoretic quantities, which serve the function of $\mathcal{I}$ (or $-\mathcal{I}$ in the case of maximisation) in our bounding algorithm: 
\begin{enumerate}
    \item[\textbf{(i)}] To measure total non-Markovianity, we employ quantum relative entropy, $S(\rho\|\sigma) := \text{Tr}\left[\rho(\log\rho - \log\sigma)\right]$, between $\Upsilon_{k:0}$ and the product of its marginals $\bigotimes_{j=1}^k\hat{\mathcal{E}}_{j:j-1} \otimes \rho_0$. This is a generalisation of {QMI} beyond the bipartite scenario and is endowed with a clear operational meaning~\cite{Pollock2018}. 
    \item[\textbf{(ii)}] To determine whether the non-Markovianity has genuine quantum features we quantify the entanglement in the process~\cite{Giarmatzi2021, milz21} by means of negativity, $\max_{\Gamma_B}\frac{1}{2}(\|\Upsilon_{k:0}^{\Gamma_B}\|_1-1)$, where $\Gamma_B$ is the partial transpose across some bipartition.\footnote{Entanglement across indices $\mathfrak{o}_j, \ \mathfrak{i}_j$ will trivially be large for nearly unitary processes since this contains information propagated by the system. We care about the entanglement across indices $\mathfrak{i}_j, \ \mathfrak{o}_{j-1}$, which measures coherent quantum memory effects between time steps as mediated by the environment.}
    \item[\textbf{(iii)}] The purity of the process, $\text{Tr} \left[ \Upsilon_{k:0}^2\right]$, measures both the strength and size of non-Markovianity through the ensemble of non-Markovian trajectories. 
    \item[\textbf{(iv)}] Finally, quantifying both Markovian and non-Markovian noise over multiple time steps, we compute the fidelity between the ideal process and the noisy process:  $\text{Tr}[\Upsilon_{k:0}(\bigotimes_{j=1}^k\ket{\Phi^+}\!\bra{\Phi^+}\otimes \rho_{\text{ideal}})]$ with $\ket{\Phi^+} = (\ket{00}+\ket{11})/\sqrt{2}$.
\end{enumerate}
Our results are summarised in Table~\ref{tab:memory-bounds}. 
We contrast these results with the coarse bounds obtained in Ref.~\cite{White-NM-2020}, where non-Markovianity is lower bounded and find the actual measures of non-Markovianity to be more than an order of magnitude higher. In the former case, maximal depolarising channels are used as causal breaks. Although this is in line with restricted process tensor capabilities, it also renders the dynamics incoherent and scrambles almost all of the information. Instead, Table~\ref{tab:memory-bounds} offers an exact estimate and is much more sensitive to the signal.

Surprisingly, we find these bounds to be extremely tight. 
This appears to be due to the relative purity of the process, with ranks (to numerical threshold of $10^{-8}$) of between 44 and 68 for the $128\times 128$ matrix {MLE} estimates, which can limit the object's free parameters~\cite{flammia2012quantum}. This is a form of compressed sensing~\cite{flammia2012quantum}, which can generally be achieved for low rank matrices. Given the relatively high coherence of the process, different unitary sequences are capable of producing highly distinguishable final states. 
When accounting for the constraints of the model (i.e., when one is restricted to the cone of process tensors), the experimental process tensor therefore appears to be resolvable without the use of mid-circuit measurements.
Process rank is also directly related to the size of the environment, with each nontrivial eigenvector constituting a relevant bath degree of freedom. In practice, this compresses requirements and estimates the effective dimension of the non-Markovian bath. We remark, however, the absence of statistical contributions to these error bars. The intention here is to demonstrate the size of bounds according to the tomographically restricted estimate. That is to say, to emphasise what can be determined by unitary sequences alone in practice. For appropriate statistical estimates, one could bootstrap error bar estimates using a binomial description of the 4096 shots collected per circuit. We estimate this not to modify any of the relevant conclusions.

Interestingly, {QMI} is much larger than zero in all situations. Moreover, from the negativity across $\mathfrak{i}_2$ and $\mathfrak{o}_1$ we find temporal entanglement in all cases. Thus, not only are these noisy stochastic processes non-Markovian, they are bipartite quantum entangled in time. This includes experiment \#1, suggesting non-Markovian behaviour is prominent beyond even the immediate effects of crosstalk. Crosstalk certainly increases correlation strength as seen in \#2, \#3, \#5, and \#6, as well as decreasing process fidelity -- an effect shown by switching neighbours into the $\ket{+}$ state and coupling to $ZZ$ noise -- but $ZZ$ noise alone cannot generate temporal entanglement~\cite{Milz2021PRXQ} (see also Appendix~\ref{app:sensor}). Note that when a DD protocol~\cite{QDD} is performed on neighbouring qubits in experiment \#4, {QMI} increases with respect to the other experiments, but with reasonably high fidelity. This may suggest that naive DD reduces noise strength while increasing complexity, warranting further studies. Lastly, the memory reduction in \#6 compared with \#2 is likely a result of information spreading away from the system (propagating to other qubits instead of back to the system). This may return to the system for a stronger memory effect, but would require a longer time scale to observe its effects.

Although this procedure is in principle to bound generic processes, in practice, we find near complete determination for only relatively few numbers of measurements. An {IC} three-step process tensor requires $12\:288$ circuits, but with only unitary gates this is $3000$, so the savings are substantial. To limit experimental overhead, one might consider increasingly measuring basis sequences until bounds achieved were satisfactorily tight in the spirit of compressed sensing. This supplements, for example, machine learning approaches to directly estimate the memory~\cite{guochu2020,Goswami2021} or full characterisations~\cite{Xiang2021}.

\section{Benchmarking Methods on Randomly Generated Processes}
\label{sec:bm}
Our work so far has looked at determining properties of process tensors via unitary operations in select scenarios. Although these scenarios constitute physically reasonable settings, we additionally would like to determine the limits of these methods in a fully general setting. To this end, we look to determining properties of \emph{random} non-Markovian processes via their unitary observables. In this setting, we are not confined to a specific model of interaction or physical environment. We construct a method by which we can tune the rank of the non-Markovian memory and sample from its distribution. The rank here is important not only because it changes the effective physics of the system, but because our unitary tools are less sensitive to higher rank processes

The generation of random states is a powerful tool used in quantum information theory to aid analysis.
Numerical methods of this sort can be used for a range of applications such as examining questions of typicality in quantum physics, testing new methods, and determining the extent to which different classes of states can possess different properties. Here, we first propose a natural ensemble of process tensors equipped with a method by which one may sample from it. Our method is constructive: generate a random state, and then project it onto the space of causal states in a way that preserves both rank and complete positivity. 
It is not our purpose to analyse the method of generation of random process in depth (although we verify some basic properties in App.~\ref{app:random}), but rather to develop a robust method that samples evenly from the flat distribution. We will then use this to look directly at properties of processes and examine the typical behaviour of restricted entanglement witnesses and process bounding.

\subsection{Random Process Tensors}

To sample random process tensors of a chosen dimension and rank, we adopt a procedure inspired by the results of Ref.~\cite{Bruzda_2009} with respect to quantum channels.
For a system of dimension $d$ over $k$ steps, let us denote the process dimension $d^{2k+1}$ by $n$ and the rank of the process tensor by $r$. The first step is to generate a state from the Ginibre ensemble. Select a matrix $X\in \mathbb{C}^{(n\times r)}$ where each element of the matrix is a normally distributed complex number to define
\begin{equation}
\rho_k = \frac{XX^\dagger}{\Tr[XX^\dagger]}.
\end{equation}
This is a unit-trace, Hermitian, PSD matrix, and hence a valid quantum state with rank $r$. But it does not satisfy causality, and hence is not a valid process. Let us begin to identify $\rho_k$ with the $k+1$-time process tensor, that is
\begin{equation}
\rho_k \in \mathscr{B}(\mathcal{H}_{\mathfrak{o}_k})\otimes \mathscr{B}(\mathcal{H}_{\mathfrak{i}_{k}}) \otimes \cdots \otimes \mathscr{B}(\mathcal{H}_{\mathfrak{o}_0}).
\end{equation}
We denote the marginals of $\rho_k$ as
\begin{equation}
\rho_{\mathfrak{i}_k,k-1} := \Tr_{\mathfrak{o}_k}[\rho_{k}]	
 \qquad \text{and} \qquad
 \rho_{k-1} := \Tr_{\mathfrak{o}_k\mathfrak{i}_k}[\rho_{k}].
\end{equation}
Now, let us define
\begin{equation}
	\breve{\rho}_{\mathfrak{i}_k,k-1} := \left(\sqrt{\rho_{\mathfrak{i}_k,k-1}}\right)^{-1} \qquad \text{and} \qquad \Lambda_{k}:=\rho_k.
\end{equation}
Since $\rho_k$ is PSD, $\rho_{\mathfrak{i}_{k},k-1}$ has a well-defined matrix square root. We will return to the existence of its inverse, but for now, assume this inverse to uniquely exist. Then let us define
\begin{equation}
\label{eq:lambda-def}
\begin{split}
	\Lambda_{k-1} &:= (\mathbb{I}_{\mathfrak{o}_k\mathfrak{i}_k} \otimes \sqrt{\rho_{k-1}})\cdot (\mathbb{I}_{\mathfrak{o}_k} \otimes \breve{\rho}_{\mathfrak{i}_k,k-1})\cdot \Lambda_{k}\cdot (\mathbb{I}_{\mathfrak{o}_k} \otimes \breve{\rho}_{\mathfrak{i}_k,k-1}) \cdot (\mathbb{I}_{\mathfrak{o}_k\mathfrak{i}_k} \otimes \sqrt{\rho_{k-1}}).
\end{split}
\end{equation}
This matrix has rank $r$. Moreover, because $A^{1/2}BA^{1/2}$ is PSD for any PSD $A,B$, we also have $\Lambda_k$ is PSD. Now we can see that if we take a partial trace over $\mathfrak{o}_k$ we obtain
\begin{equation}
\label{eq:lambda-causal}
\Tr_{\mathfrak{o}_k}[\Lambda_{k-1}] = (\mathbb{I}_{\mathfrak{i}_k}\otimes \sqrt{\rho_{k-1}})\cdot \breve{\rho}_{\mathfrak{i}_k,k-1} \cdot \rho_{\mathfrak{i}_{k}, k-1} \cdot \breve{\rho}_{\mathfrak{i}_k,k-1} \cdot (\mathbb{I}_{\mathfrak{i}_k}\otimes \sqrt{\rho_{k-1}})
=\mathbb{I}_{\mathfrak{i}_k}\otimes\rho_{k-1}.
\end{equation}

$\Lambda_{k-1}$ is hence locally causal, in that tracing over the final output reduces to an identity on its input leg, but the remaining marginal $\rho_{k-1}$ does not iteratively satisfy the same conditions. 
Next, we define $\Lambda_{k-2}$ as
\begin{equation}
\Lambda_{k-2} := (\mathbb{I}_{\mathfrak{i}_k,\mathfrak{i}_k-1} \otimes \sqrt{\rho_{k-2}})\cdot (\mathbb{I}_{\mathfrak{i}_k,\mathfrak{o}_{k-1}} \otimes \breve{\rho}_{\mathfrak{i}_{k-1},k-2})\cdot \Lambda_{k-1} \cdot (\mathbb{I}_{\mathfrak{i}_k,\mathfrak{o}_{k-1}} \otimes \breve{\rho}_{\mathfrak{i}_{k-1},k-2}) \cdot (\mathbb{I}_{k,k-1} \otimes \sqrt{\rho_{k-2}})
\end{equation}
to get the next causal condition: $\text{Tr}_{\mathfrak{o}_k,\mathfrak{o}_{k-1}}[\Lambda_{k-1}] = \mathbb{I}_{\mathfrak{i}_k,\mathfrak{i}_{k-1}} \otimes \rho_{k-2}$. We can iterate our process by setting, for $0\leq j \leq k-1$:
\begin{equation}
\begin{split}
	\Lambda_j &:= (\mathbb{I}_{k,\mathfrak{o}_{k-1}\mathfrak{i}_k \dots \mathfrak{o}_j\mathfrak{i}_j} \otimes \sqrt{\rho_{j}})  \cdot (\mathbb{I}_{\mathfrak{o}_k\mathfrak{i}_k \dots \mathfrak{o}_j} \otimes \breve{\rho}_{\mathfrak{i}_{j+1},j})\cdot \Lambda_{j+1} \cdot (\mathbb{I}_{\mathfrak{o}_k\mathfrak{i}_k \dots \mathfrak{o}_j} \otimes \breve{\rho}_{\mathfrak{i}_{j+1},j}) \cdot (\mathbb{I}_{\mathfrak{o}_k\mathfrak{i}_k \dots \mathfrak{o}_j\mathfrak{i}_j} \otimes \sqrt{\rho_{j}}).
\end{split}
\end{equation}
The subsequent transformations do not affect the relation in Eq.~\eqref{eq:lambda-causal}, and hence the end result is $\Lambda_0$, a random rank $r$ process tensor of a $k$-step process.

The above derivation presupposed the existence of the matrix inverse of $\rho_{\mathfrak{i}_k,k-1}$. This matrix is defined through the marginalisation of $\rho_k$ over $\mathfrak{o}_k$, which has size $d$. This means $\text{rank}(\rho_{\mathfrak{i}_k,k-1}) = \text{min}(r d_S, n/d_S)$. Thus, if $r < n/d_S^2$, then $\rho_{\mathfrak{i}_k,k-1}$ is singular and hence cannot be inverted. This is problematic if we want to examine low-rank processes. To resolve this, we consider an `inner' iteration of the algorithm.
Let us instead set 
\begin{equation}
	\breve{\rho}_{\mathfrak{i}_{k},k-1} = (\sqrt{\rho_{\mathfrak{i}_{k},k-1}})^{+},
\end{equation}
where $(\cdot)^+$ denotes the Moore-Penrose pseudoinverse. This transformation takes the inverse on the image (Im) of $\rho_{\mathfrak{i}_{k},k-1}$, but leaves the kernel (Ker) untouched. The process is locally causal on the support of the marginal $\rho_{\mathfrak{i}_{k},k-1}$. Consider its eigendecomposition:
\begin{equation}
	\rho_{\mathfrak{i}_{k},k-1} = \sum_{i=1}^{r < n/d_S^2}p_i|\psi_i\rangle\!\langle \psi_i|,
\end{equation}
where $\{|\psi_i\rangle\}$ is an orthonormal basis of $\text{Im}(\rho_{\mathfrak{i}_{k},k-1})$. Let $\{|\phi_i\rangle\}$ be an orthonormal basis of $\text{Ker}(\rho_{\mathfrak{i}_{k},k-1})$
we have
\begin{equation}
\breve{\rho}_{\mathfrak{i}_{k},k-1} = \sum_{i=1}^{r} \frac{1}{\sqrt{p_i}}|\psi_i\rangle\!\langle\psi_i|.
\end{equation}
We see then that 
\begin{equation}
	\label{eq:deficient-projection}
  \breve{\rho}_{\mathfrak{i}_{k},k-1} \cdot \rho_{\mathfrak{i}_{k},k-1} \cdot \breve{\rho}_{\mathfrak{i}_{k},k-1}
= \sum_i |\psi_i\rangle\!\langle \psi_i| = \mathbb{I}_{\mathfrak{i}_{k},k-1} - \sum_{j=1}^{d_S^{2k}-r}|\phi_j\rangle\!\langle \phi_j|.
\end{equation}

The process is hence causal on the support of $\rho_{\mathfrak{i}_k,k-1}$. Let $\Lambda_k^{(0)}:= \rho_k$ and similarly for $\rho_{\mathfrak{i}_k,k-1}^{(0)}$ and $\breve{\rho}_{\mathfrak{i}_k,k-1}^{(0)}$. Then recursively define 
\begin{equation}
    \Lambda_{k}^{(l+1)} := (\mathbb{I}_{\mathfrak{o}_k\mathfrak{i}_k}^{(l)} \otimes \sqrt{\rho_{k-1}^{(l)}})\cdot (\mathbb{I}_{\mathfrak{o}_k} \otimes \breve{\rho}_{\mathfrak{i}_k,k-1}^{(l)})\cdot \Lambda_{k}^{(l)} \cdot (\mathbb{I}_{\mathfrak{o}_k} \otimes \breve{\rho}_{\mathfrak{i}_k,k-1}^{(l)}) \cdot (\mathbb{I}_{\mathfrak{o}_k\mathfrak{i}_k}^{(l)} \otimes \sqrt{\rho_{k-1}^{(l)}}).
\end{equation}
At each iteration, the spectrum of $\Lambda_{k-1}^{(l+1)}$ will be different to the last (starting from a random state), and its rank will be unchanged. Moreover, because $\rho_{k-1}^{(l)}$ does not act on the $\mathfrak{i}_k$ subsystem, the state will remain causal on $\bigcup_{l}\text{Im}(\rho_{\mathfrak{i}_k,k-1}^{(l)})$. Combining these two facts, we have that eventually the algorithm converges such that
\begin{equation}
    \text{Tr}_{\mathfrak{o}_k}\left[\Lambda_{k}^{(l)}\right] = \mathbb{I}_{\mathfrak{i}_k}\otimes \Tr_{\mathfrak{o}_k\mathfrak{i}_k}[\Lambda_k^{(l)}]
\end{equation}
where $\Lambda_k$ still has rank $r$ for all $l$.

The number of iterations is at worst $d^{2k}-r$, but in practice we find fewer than this are required. Note, however, that if $r\ll d$ the computation may become numerically unstable. This process is summarised in Algorithm~\ref{alg:random-pts}.

\begin{algorithm}[t]
\caption{Generating Random Process Tensors}
\label{alg:random-pts}
\begin{algorithmic}[1]
\Input{System dimension $d$, no. steps $k$, rank $r$, maximum iterations $M$.}
\Output{Random $k$-step process tensor $\Upsilon_{k:0}$ with environment size $r$.}
\State $j\gets0,$ $n \gets {d_S}^{2k+1}$
\State $X\gets$ \text{RandomGinibreMatrix($n,r$)}
\State $\rho \gets {XX^\dagger}/{\text{Tr}[XX^\dagger]}$ \Comment{Start with a quantum state.}
\State $\Upsilon \gets \rho$
\State $\Upsilon_{\rm c}$, $\Upsilon_{\rm r}$ \Comment{Initialise cancellation and remainder terms.}
\While{$\Upsilon \not\subset \mathcal{V}_{\rm c}$ \text{ and } $j\leq M$}
\State $\Upsilon = \Upsilon / \text{Tr}[\Upsilon]$
    \For{$i \gets 1 \text{ to } k$} \Comment{Impose causality for each time.}
        \State $\Upsilon_{\rm c} = \text{Tr}_{[1:2i+1]}[\Upsilon]$
        \State $\Upsilon_{\rm r} = \text{Tr}_{[1:2i+2]}[\Upsilon]$
        \State $\Upsilon_{\rm c} = (\sqrt{\Upsilon_{\rm c}})^{+}$
        \State $\Upsilon_{\rm r} = \sqrt{\Upsilon_{\rm r}}$
        \State $\Upsilon = (\mathbb{I}_{2i+1}\otimes \Upsilon_{\rm c})\cdot \Upsilon \cdot (\mathbb{I}_{2i+1}\otimes\Upsilon_{\rm c})$
        \State $\Upsilon = (\mathbb{I}_{2i+2}\otimes \Upsilon_{\rm r})\cdot \Upsilon \cdot (\mathbb{I}_{2i+2}\otimes\Upsilon_{\rm r})$ \Comment{Positive projection onto set of causal processes.}
        \State $ j = j+1$
    \EndFor
\EndWhile
\State\Return $\Upsilon_{k:0} \gets \Upsilon$
\end{algorithmic}
\end{algorithm}

\subsection{Sampling Typical Process Tensors}

\begin{figure}[!t]
	\centering
	\includegraphics[width=\linewidth]{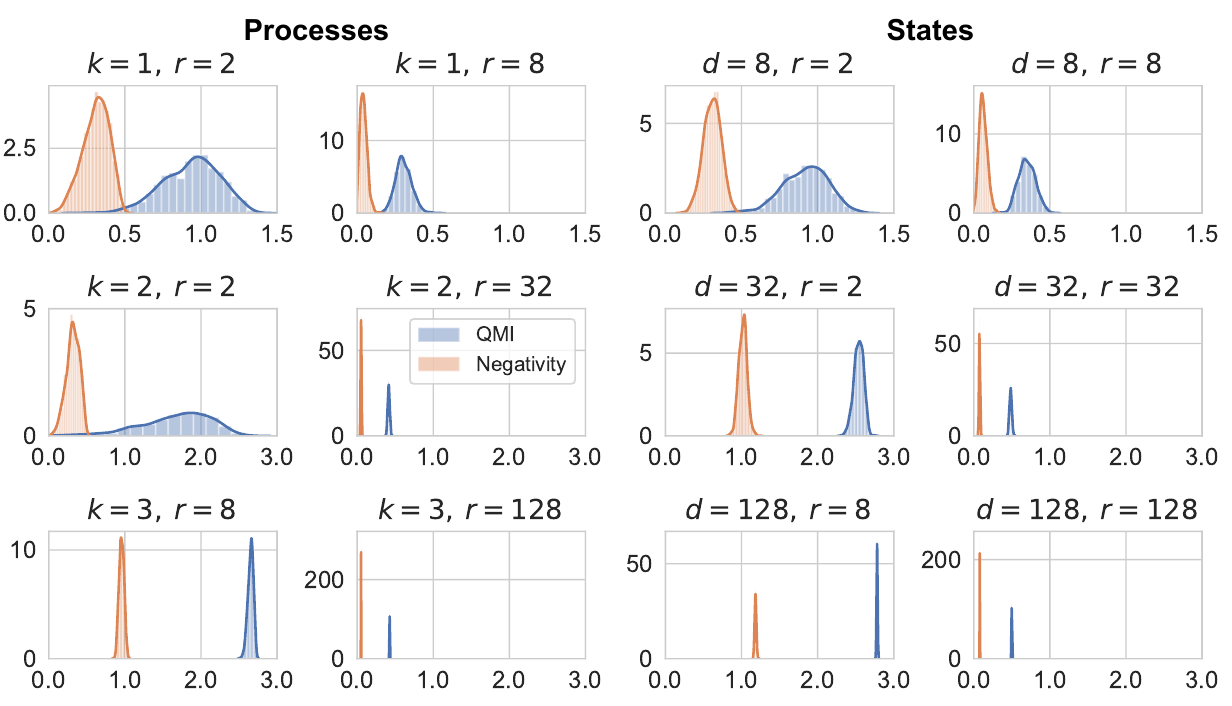}
	\caption[Correlation histograms for randomly generated process tensors of high and low rank ]{Histograms of total correlation and bipartite entanglement for 1000 randomly sampled processes and states of high and low rank. We see the same general trend, that typical processes can have highly complex temporal correlations much like typical states. However we note the widening of these distributions, and lower typical mean. This is most clearly seen when comparing the $d=32$, $r = 2$ cases for states and processes.}
	\label{fig:process-properties}
\end{figure}

To begin we will showcase some numerics that demonstrate some properties of typical process tensors. In addition to the technique itself, what we wish to impress is that multi-time processes can be as complex as many-body quantum states. Figure~\ref{fig:process-properties} shows a series of histograms displaying the populations of different properties of both states and processes. For $k=1,2,$ and $3$, we draw random low-rank and high-rank process tensors, as well as equivalent dimension states. We compute the generalised quantum mutual information $\mathcal{N}(\Upsilon_{k:0})$,
as well as negativity across the respective bipartitions 
\begin{equation}
	\begin{split}
		&\{\{\mathfrak{o}_1,\mathfrak{i}_1\}, \{\mathfrak{o}_0\}\},\\
		&\{\{\mathfrak{o}_2,\mathfrak{i}_2\}, \{\mathfrak{o}_1,\mathfrak{i}_1,\mathfrak{o}_0\}\},\quad\text{and}\\
		&\{\{\mathfrak{o}_3,\mathfrak{i}_3,\mathfrak{o}_2,\mathfrak{i}_2\}, \{\mathfrak{o}_1,\mathfrak{i}_1,\mathfrak{o}_0\}\}.
	\end{split}
\end{equation}
We observe several key features. First, note that in the typical cases, both quantum and total correlation between states and processes are commensurate. 
Causality constraints do not significantly limit the quantum properties that a process may exhibit.

\begin{figure}[t!]
	\centering
	\includegraphics[width=\linewidth]{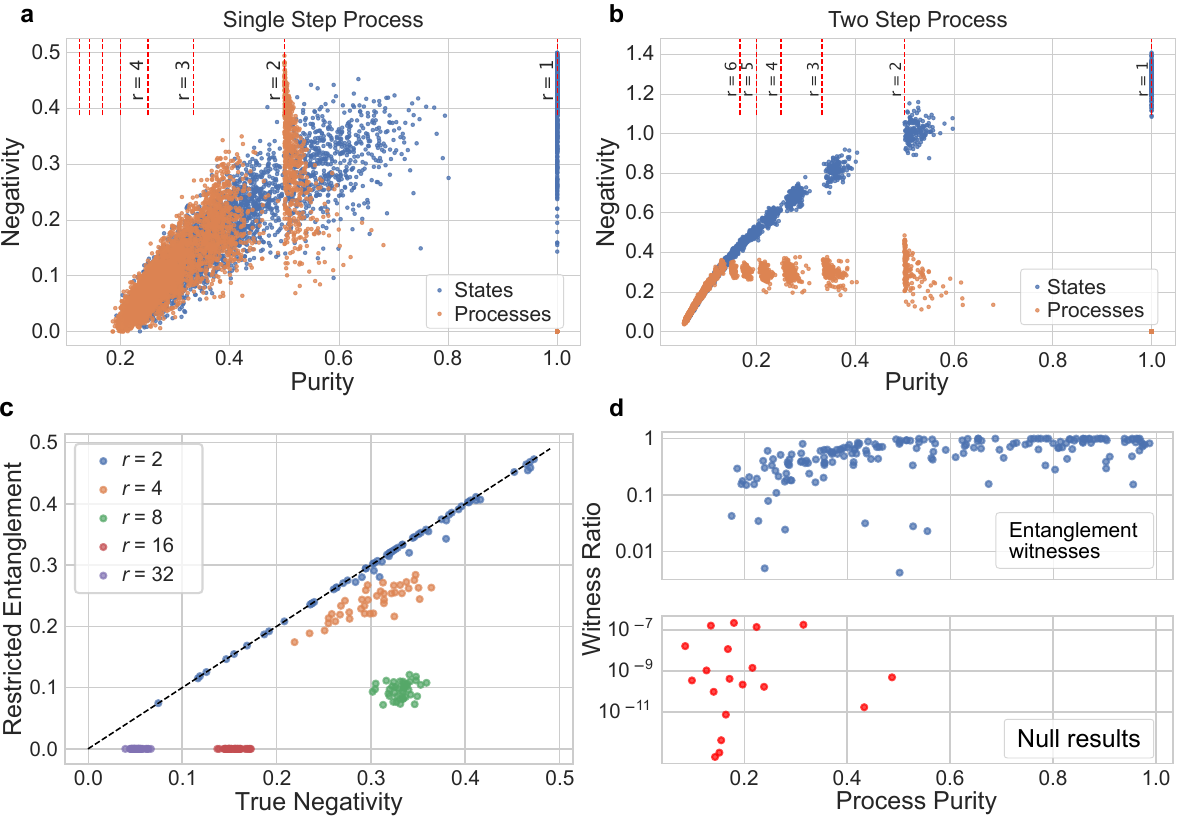}
	\caption[Restricted entanglement witnessing for randomly generated single and two-step process tensors ]{Results of restricted entanglement witnessing for randomly generated two-step process tensors. Relationship between purity and negativity in randomly generated states and processes at different ranks, \textbf{a} for $k=1,d=8$ and \textbf{b} for $k=2,d=32$. We see the effects of causality more clearly here. Typical processes do not follow the same purity-negativity trend as typical states: the features are more tightly distributed, and tend to zero where $\gamma = 1$. \textbf{c} We compare the true negativity of a given process with the negativity as bounded by unitary observables, generated across different ranks. \textbf{d} We plot the ratio of witnesses as a function of the purity of different processes. The top panel shows all the instances where entanglement is sufficiently measured by unitary observables. The bottom panel shows that for highly mixed processes, we get some instances where entanglement cannot be observed in the restricted regime. }
	\label{fig:entanglement-benchmarking}
\end{figure}

It is also instructive to observe the general trend of how entanglement behaves as a function of purity in typical cases. As a general rule, the more pure a random quantum state is, the more entangled it is. However, a maximally pure process is necessarily a Markovian one for fixed system dimension. Rather, then, than total correlation scaling roughly as a function of purity, processes maximise their correlation at an intermediate purity -- where the environment is not too large to scramble, but not so pure as to lack an environment that mediates correlations at all. We show numerically the relationship between temporal entanglement and purity of a process for single and two-step process tensors in Figures~\ref{fig:entanglement-benchmarking}a and b. Note that a state with rank $r$ always has purity at least $1/r$ (saturated for an equal mixture of $r$ pure states).

\subsection{Detecting Entanglement}

Using these methods to draw random process tensors, let us now benchmark our restricted entanglement witnesses across the ensemble.
We first compute the negativity across $\{\{\mathfrak{o}_2,\mathfrak{i}_2\},\{\mathfrak{o}_1,\mathfrak{i}_1,\mathfrak{o}_0\}\}$, and then solve the restricted {SDP} from Section~\ref{sec:temp-entanglement} to determine an entanglement bound as certified by the set of restricted observables. In Figure~\ref{fig:entanglement-benchmarking}a, we plot the restricted entanglement bound against the true negativity for random processes generated with increasing ranks, denoted by $r$. Interestingly, we see that up to rank eight process, restricted observables always suffice to witness temporal entanglement, and for rank two processes the resulting amount is almost identical to the true negativity. This can be seen that when the process is more mixed, it is harder to extract a signal with unitary operations without getting drowned out in the noise. Whereas a causal break purifies the process for any particular run, allowing correlations to be detected.

This effect is shown plotted in Figure~\ref{fig:entanglement-benchmarking}b, where we have generated processes via random Heisenberg interactions with an $n$-qubit environment $1\leq n \leq 5$. This produces a range of processes with purities that depend on the strength of the interaction. We plot the ratio between the unitary-witnessed entanglement, and the true entanglement value. For very low purities, unitary observables often fail to detect the entanglement, but it is significantly more reliable for even relatively pure processes.

\subsection{Bounding Mutual Information}

\begin{figure}[t]
	\centering
	\includegraphics*[width=0.7\linewidth]{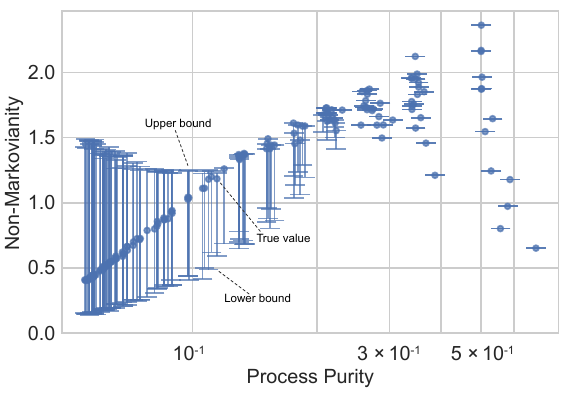}
	\caption[Bounds on generalised quantum mutual information for random restricted process tensors ]{Bounds given on generalised {QMI} for randomly generated two-step process tensors, plotted against purity. For mixed processes, we see that the bounds are sufficient to detect non-Markovian correlations, but are reasonably wide. With increasing purity, the bounds tighten until unitary data appears sufficient to fully determine the memory of a process.}
	\label{fig:random-PT-bounds}
\end{figure}

We can additionally evaluate the performance of generalised {QMI} bounds for random process tensors, as per the method introduced in Section~\ref{sec:bounding}.
We saw in Table~\ref{tab:memory-bounds} that memory bounds obtained on real data in practice are tight. It is natural, then, to wonder if this is a generic property of processes and how the bounds behave for typical instances. To investigate this, we apply our bounding procedure for randomly generated process tensors, shown in Figure~\ref{fig:random-PT-bounds}. Here, we generate two-step processes and determine their non-Markovianity in the form of generalised {QMI}. We then optimise to find lower and upper bounds for the memory that are consistent with the set of restricted observables. We plot this against the purity of the generated process.
For processes of low purity, the bounds are wide, but well above zero. For higher purity processes, however, the bounds tighten significantly. This is consistent with the set of observations so far that unitary operations in practice suffice to completely determine the process when the dynamics are not too mixed. Heuristically, then, unitary dynamics appears to be sufficient to uniquely determine low-rank processes.
It would be an interesting future research direction to place guarantees on the size of the bounds that may be obtained from restricted processes. 

Figures~\ref{fig:entanglement-benchmarking} and~\ref{fig:random-PT-bounds} indicate that, although we can learn a surprising amount of information about multi-time correlations via sequences of unitaries, it is only true in the case of relatively pure or coherent processes. For more rigorous guarantees in the context of mixed processes, one will always have to turn to an informationally complete set of controls. Intuitively, this can be understood that the more mixed a process is, the less sensitive the final measurement will be to any early unitary operations. With access to measurements at all times, however, one can purify the process and extract these correlations directly. 

\section{Discussion}
\label{sec:dis}
Although the study of many-body physics has uncovered a rich tapestry of structure in different corners of applied quantum mechanics, temporal quantum correlations remain relatively under-explored. In this work, we have motivated, demonstrated, and made accessible the study of many-time physics beyond the two-time correlators regime. We have performed this investigation under the practical lens where only unitary control is afforded. Not only is this a conceptually interesting standpoint -- considering that unitaries are deterministic operations -- but it is of practical importance, given that measurement and feedback is often a slow, invasive, and noisy process.
Multi-time correlations will showcase the full potential of the field of many-time physics, including capturing dynamical phases of matter~\cite{heyl2018dynamical}.

A primary achievement is in showing how rich non-Markovian effects can be studied with only unitary control and a final projective measurement. Rather than merely simulating non-Markovian quantum processes on fully controllable quantum computers, this allows observation of complex naturally occurring phenomena~\cite{PhysRevA.61.023603,jaksch2005cold,PhysRevLett.101.260404,PhysRevA.90.032106,caruso2009highly,lambert2013quantum}. One might imagine, for example, the use of quantum sensors in condensed matter or biological systems. Nitrogen-vacancy centres in diamond, for example, can be unitarily controlled with arbitrary waveform generators, projectively read out optically, and have been shown to be biocompatible~\cite{McGuinness2011}. The present work helps to elucidate the role of quantum mechanics in open quantum systems without needing to understand the complete system. It can therefore a powerful supplementary tool for quantum sensors that wish to understand an unknown environment.\par 

On the more applied side of things, our results on IBM Quantum devices show that even idle qubits can exhibit surprisingly complex and temporally correlated dynamics. We have discovered that the non-Markovianity persists not simply as a classical set of correlations, but as temporal quantum entanglement. This prompts the need for further study of non-Markovian behaviour on {NISQ} devices. Specifically, if not fabricated away, correlated noise ought to be able to be converted into clean channels through appropriate control sequences~\cite{berk2021extracting}.

\begin{acknowledgments}
We sincerely thank the anonymous referees for their comprehensive feedback and numerous useful suggestions.
This work was supported by the University of Melbourne through the establishment of an IBM Quantum Network Hub at the University.
G.A.L.W. is supported by an Australian Government Research Training Program Scholarship. 
C.D.H. is supported through a Laby Foundation grant at The University of Melbourne. 
K.M. is supported through Australian Research Council Future Fellowship FT160100073.
K.M. and C.D.H. acknowledge the support of Australian Research Council's Discovery Project DP210100597.
K.M. and C.D.H. were recipients of the International Quantum U Tech Accelerator award by the US Air Force Research Laboratory.
\end{acknowledgments}
%

\appendix

\section{Classically Correlated Processes}\label{app:sensor}
In this appendix, we define and describe some essential features of classically correlated quantum processes. \textit{Classical} here, means that the memory can be described solely by a classical probability distribution, and could be replicated in practice with only classical resources. We connect this description to the process tensor, and show that the class of dynamics from locally diagonalisable Hamiltonians constitutes a large portion of these processes. 
By the state-process equivalence, processes can carry non-trivial physics through correlations in time as mediated by environmental memory. In considering process tensor properties, we are interested in establishing a boundary of sorts between simple processes and more complicated ones. Specifically, an important question we will come to in the study of correlated noise is whether the noise can be categorised as classical or quantum. By classical, we mean here not necessarily that the process only interacts with classical fields in its dynamics. Indeed, we will see that completely quantum environments might be called classical. What we mean is that the hypothetical quantum memory could be entirely simulated classically. That is, the correlations can be entirely explained by classical random variables. This constitutes a set of dynamics without any quantumness in its temporal correlations. This concept is illustrated as follows:
\begin{definition}
	If a process tensor $\Upsilon_{k:0}$ can be expressed in terms of quantum trajectories
	\begin{equation}
		\label{eq:p-sep}
		\Upsilon_{k:0} = \sum_{i=1}^{2^{2k}} p_i \left(\bigotimes_{j=1}^k \hat{\mathcal{E}}_{j:j-1}^{(i)} \otimes \rho_0^{(i)}\right),
	\end{equation}
	where each $\hat{\mathcal{E}}_{j:j-1}^{(i)}$ is a valid CPTP map and each $p_i\geq 0$, then this is said to be \emph{process separable}, and the correlations are classically simulable.
\end{definition}
\begin{figure}
	\centering
	\includegraphics[width=0.3\linewidth]{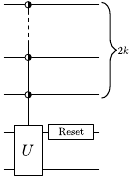}
	\caption[Circuit depiction of a classically correlated process. ]{Circuit depiction of a classically correlated process. }
	\label{fig:p-sep-multiplex}
\end{figure}
To unpack this slightly, we choose the terminology to be process separable rather than just separable because the Choi state is not completely separable -- that is, entanglement can exist between $\mathfrak{o}_j,\mathfrak{i}_j$ pairs. Entanglement across $\mathfrak{o}_j,\mathfrak{i}_j$ simply communicates the fact that process marginals do not destroy coherence of the incoming states, but that there is no non-Markovian temporal entanglement~\cite{zambon2023relations}. We say that the correlations are classically simulable because no quantum resources are required in the environment. Although Eq.~\eqref{eq:p-sep} is non-Markovian, it is a convex combination of Markovian processes. Thus, it fits into the class of dynamics known as quantum trajectories because the dynamics can essentially be simulated as a convex combination of system control protocols. On a shot-by-shot basis, one could draw an item $i_0$ from the classical distribution $\{p_i\}$, and then implement the dynamics $\bigotimes_{j=1}^k\hat{\mathcal{E}}_{j:j-1}^{(i_0)}\otimes \rho_0^{(i_0)}$. \par 

A useful way to think about processes of this type is that we can always dilate them to an environment implementing multiplex gates. Multiplex gates are generalisations of controlled-unitaries that implement unitaries on a target dependent on the bit string of the control register. For example, a multiplex gate with two control qubits can be written in block-diagonal form:
\begin{equation}
	\begin{pmatrix} 
		u_{00} & & & \\
		& u_{01} & & \\
		& & u_{10} & \\
		& & & u_{11}
	\end{pmatrix},
\end{equation}
where the unitary $u_{ij}$ is applied to the target qubits conditioned on the control qubits being in state $|ij\rangle$. Now, any process tensor in the form of Eq.~\eqref{eq:p-sep} can be written with $2k$ environment qubits and $2$ ancilla qubits. In each step, $2k$ of them are controls of a multiplexed gate targeting an $SU(4)$ gate on the ancillas and the system. The ancillas are then reset to any pure state. This is depicted in Figure~\ref{fig:p-sep-multiplex}. The purpose of the ancillas is to allow the possibility that the $\hat{\mathcal{E}}_{j:j-1}^{(i)}$ be non-unital, and the reset to ensure that no quantum memory is carried forward between each step. Let us first eliminate the ancilla qubits, and have the target unitaries act only on the system. Note that as a result, the individual $\hat{\mathcal{E}}_{j:j-1}^{(i)}$ are mixed unitary channels. Let us denote a multiplex gate between times $t_{j-1}$ and $t_j$ by $M_{j:j-1}$.


We can relate this to a particular class of dynamics by converting processes of this mixed unitary type to a Hamiltonian model. Consider first a restriction of the controlled unitary to being a rotation about $Z$, $R_Z$.

\begin{figure}
	\centering
	\includegraphics[width=0.3\linewidth]{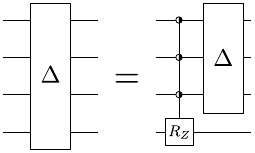}
	\caption[Decomposition of a diagonal gate into a multiplexed $Z$-rotation and another diagonal gate. ]{Decomposition of a diagonal gate into a multiplexed $Z$-rotation and another diagonal gate. }
	\label{fig:diag-to-rz}
\end{figure}
We make use of the fact that one may decompose a diagonal gate $\Delta$ on $n$ qubits to the composition into a composition of a multiplexed $R_Z$ on $n$ qubits followed by a diagonal gate (in the $Z$-basis) on the $n-1$ target qubits, as in Figure~\ref{fig:diag-to-rz}. 
The remaining $\Delta$ on $E$ commutes with any subsequent multiplex and can be brought to the end of the circuit and removed by the partial trace. Hence, any diagonal $SE$ interaction generates a process tensor of the form in Eq.~\eqref{eq:p-sep}. We can add more structure to this by introducing the following definition from Ref.~\cite{cubitt2016complexity}.
\begin{definition}
	Suppose we have a Hamiltonian $H_{SE}\in \mathcal{B}(\mathcal{H}_{S}\otimes \mathcal{H}_E)$. If 
	\begin{equation}
		H_{SE} = (u_S\otimes u_E) D_{SE} (u_S^\dagger\otimes u_E^\dagger),
	\end{equation}
	for diagonal $D_{SE}$, we say the Hamiltonian is \emph{locally diagonalisable}.
\end{definition}
It is easy to check that this definition implies
\begin{equation}
	\label{eq:local-H-structure}
	H_{SE} = \alpha A\otimes B + \beta A\otimes \mathbb{I} + \gamma \mathbb{I} \otimes B + \delta \mathbb{I}\otimes \mathbb{I},
\end{equation}
for real $\alpha, \beta, \gamma, \delta$, and Hermitian $A\in \mathcal{B}(\mathcal{H}_S), B\in \mathcal{B}(\mathcal{H}_E)$. We can show that dynamics generated by locally diagonalisable Hamiltonians also produce process-separable process tensors by noting that:
\begin{equation}
	\begin{split}
		U_{j:j-1}^{SE} &= \exp\left(-i(t_j - t_{j-1}) H_{SE}\right)\\
		&= (u_S\otimes u_E) \exp \left(-1 (t_j - t_{j-1})D_{SE}\right) (u_S^\dagger\otimes u_E^\dagger)\\
		&= (u_S\otimes u_E) \Delta_{j:j-1} (u_S^\dagger\otimes u_E^\dagger)\\
		&= (u_S\otimes u_E) M_{j:j-1}(R_Z) (u_S^\dagger\otimes u_E^\dagger)
	\end{split}
\end{equation}
Hence, for sequential $U_{SE}$:
\begin{equation}
	\begin{split}
		U_{j+1:j} \cdot U_{j:j-1} &= (u_S\otimes u_E) \Delta_{j+1:j} \Delta_{j:j-1} (u_S^\dagger\otimes u_E^\dagger)\\
		&= U_{j+1:j} \cdot U_{j:j-1} = (u_S\otimes u_E) M_{j+1:j}(R_Z) M_{j:j-1}(R_Z)(u_S^\dagger\otimes u_E^\dagger),
	\end{split}
\end{equation}
The $u_S$ applies a similarity transformation to each $R_Z$ in the multiplex. Meanwhile, each $u_E$ cancels out except for the very first and last. The first can be absorbed into the definition of the environment state, and the last absorbed into the partial trace. Hence, a concatenation of $k$ $SE$ evolutions generated by locally diagonalisable Hamiltonians produces a process tensor which is process-separable, as in Eq.~\eqref{eq:p-sep}. Note also that any operation applied to $S$ does not change this structure, and so we can have the system-only part of $H_{SE}$ be arbitrary too, i.e. $H_S = \gamma I_E\otimes C$ for some Hermitian $C$. 

This provides us some insight into the complexity of different open quantum systems. Specifically, we can draw a distinction between non-Markovian processes whose memory can be simulated with classical resources only, versus those that need some quantumness. An interesting consequence of this is that $SE$ interactions that generate entanglement between a system and its environment is not sufficient to generate temporal entanglement, temporally entangled processes constitute a more restrictive class than the state equivalence. Below are some examples of $SE$ interactions (independent of environment state) that generate classically correlated processes:

\begin{itemize}
	\item Quantum Ising models $H_{SE} = \sum_{ij\in \mathcal{C}} J_{ij} Z_{i}\otimes Z_{j} + \sum_i Z_i$, modelling for example $ZZ$ coupling found in transmon devices with connectivity graph $\mathcal{C}$.
	\item A sequence of CNOT gates between system and environment, by virtue of the fact that CNOTs can be generated by a Hamiltonian $H_{CNOT}\propto (\mathbb{I} - X)\otimes (\mathbb{I} - Z)$. 
	\item Kitaev's toric code Hamiltonian~\cite{KITAEV20032}, with respect to any given qubit on a lattice: $H = -\sum_{\nu} A(\nu) - \sum_p B(p)$, where $A(\nu) = \sigma_{\nu,1}^x\sigma_{\nu,2}^x\sigma_{\nu,3}^x\sigma_{\nu ,4}^x$ and $B(p) = \sigma_{p,1}^z\sigma_{p,2}^z\sigma_{p,3}^z\sigma_{p,4}^z$. Here, $\nu$ are indices of a vertex on the lattice and $p$ are indices of a plaquette. $A$ and $B$ terms always commute since vertices and plaquettes share an even number of sites, hence the interaction is locally diagonalisable and leads only to classical temporal correlations. For clarity here, the system is a single qubit with respect to the rest of the array as an environment. 
\end{itemize}

In the main text, $H_I^{\text{(Classical)}} = Z\otimes(X+Z)$ takes this locally diagonalisable form and is hence responsible for classical memory only.

\section{Spectral Properties of Random Processes}
\label{app:random}

Here, we briefly analyse the spectral properties of randomly generated process tensors, illustrating conventional features of randomness in our sampling algorithm. This structure can be better seen by transforming the process tensor Choi state $\Upsilon_{k:0}$ to superoperator form, $\Phi_{k:0}$. 
\begin{figure}[!b]
	\centering
	\includegraphics[width=\linewidth]{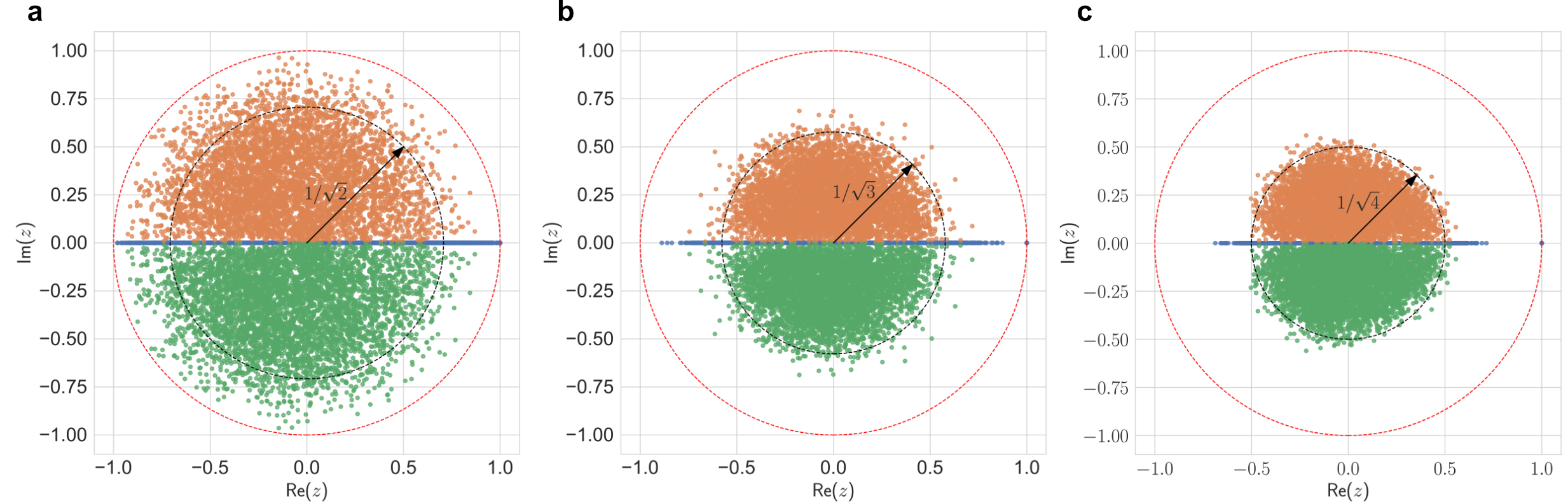}
	\caption[Eigenvalue distribution of 5000 random two-step process tensors ]{Eigenvalue distribution of 5000 random two-step process tensors. Plotted alongside the unit disc and the Girko disc $1/\sqrt{r}$. \textbf{a} $r=2$, \textbf{b} $r = 3$, and \textbf{c} $r = 4$. Orange, blue, and green colouring are eigenvalues with positive, zero, and negative imaginary components, respectively.}
	\label{fig:eval_dist}
\end{figure}
Since process tensors are CP, then a given process tensor $\mathcal{T}_{k:0}$ has operator-sum decomposition,
\begin{equation}
	\rho[\mathbf{A}_{k-1:0}] = \mathcal{T}_{k:0}[\mathbf{A}_{k-1:0}] = \sum_{l=1}^r T_l \hat{\mathbf{A}}_{k-1:0} T_l^\dagger.
\end{equation}
The Choi state is then defined as 
\begin{equation}
	\Upsilon_{k:0} = \sum_{l=1}^r |T_l\rangle\!\rangle \langle \!\langle T_l|,
\end{equation}
and superoperator form consequently as 
\begin{equation}
	\Phi_{k:0} = \sum_{l=1}^r T_l^\ast\otimes T_l.
\end{equation}
This is equivalent to a reshuffling operation $(\cdot)^R$,
\begin{equation}
	(|k_{\mathfrak{o}_j}\rangle\!\langle b_{\mathfrak{o}_j}| \otimes |k_{\mathfrak{i}_j}\rangle\!\langle b_{\mathfrak{i}_j}|)^R = | k_{\mathfrak{o}_j}\rangle\!\langle k_{\mathfrak{i}_j}| \otimes |b_{\mathfrak{o}_j} \rangle\!\langle b_{\mathfrak{i}_j}| \:\forall\: j,
\end{equation}
such that $\Phi_{k:0} = \Upsilon_{k:0}^R$. For quantum channels $\mathcal{E}$, the spectrum is contained within the unit disc $\{z\in \mathbb{C} : |z|\leq 1\}$, and there is a leading eigenvalue $\lambda_1 = 1$~\cite{Kukulski_2021}. The spectrum is also symmetric about the real axis. We can view process tensors as a CPTP map from the space of inputs in a multi-time process to the space of outputs. As a consequence, its spectrum obeys the same conditions. In the limit of large dimension, random quantum channels drawn from a uniform measure have the same first two cumulants as the Gaussian distribution, with negligible higher order contributions. The trailing eigenvalues therefore tend to concentrate in the so-called \emph{Girko disc} of $1/\sqrt{r}$, recalling that $r$ is the rank, or number of Kraus operators. We see this too with process tensors. Expressed in superoperator form, Figure~\ref{fig:eval_dist} shows both the leading eigenvalues in the unit disc, the symmetry about $\text{Re}(z)$, and the clustering of eigenvalues in the Girko disc. From this we see good evidence that our method to sample processes is indeed according to the flat distribution.

\section{Experimental Details}
\label{app:exp}

\begin{figure}[htbp]
	\centering
	\includegraphics[width=0.5\linewidth]{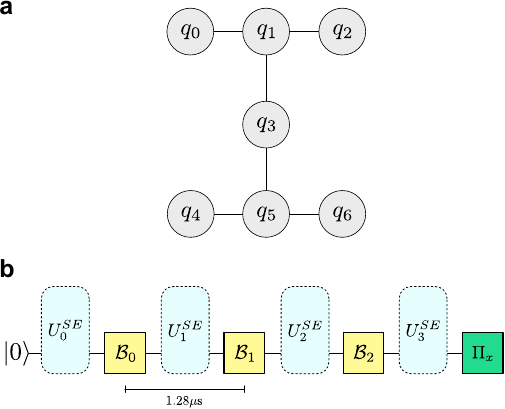}
    \caption[Experiments conducted on \emph{ibmq\_casablanca} to determine non-Markovian memory ]{\textbf{a} Qubit map of \emph{ibmq\_casablanca}, $q_5$ was the system qubit in all experiments. \textbf{b} Structure of the process tensor circuits for this device: a three-step process tensor constitutes three basis operations, followed by measurement. For all six experiments, we fixed the wait time at $1.28\:\mu$s and varied the background.}
	\label{fig:casablanca_layout}
\end{figure}

The layout of the device from Table~\ref{tab:memory-bounds} is shown in Fig~\ref{fig:casablanca_layout}a and the structure of the process tensor experiments in Fig~\ref{fig:casablanca_layout}b. $q_5$ was used as the system qubit for all experiments shown in Table~\ref{tab:memory-bounds}, and had a measured $T_2$ time of 123.65 $\mu$s during the period over which data was collected. For experiments \#1 to \#6 the backgrounds were varied, and the wait time between each time step identically fixed to 1.28 $\mu$s. A ten-unitary basis was used, for a total of $3000$ circuits per experiment at $4096$ shots. The details of each background are as follows: 
\begin{enumerate}
    \item No control operations on any other qubits.
    \item $q_3$, $q_4$, and $q_6$ were each initialised into a $\ket{+}$ state at the beginning of the circuits and left idle.
    \item $q_3$, $q_4$, and $q_6$ were each initialised into a $\ket{+}$ state at the beginning of the circuits, and, in each wait period, four sequential CNOTs controlled by $q_3$ and with $q_1$ as a target were implemented. 
    \item $q_3$, $q_4$, and $q_6$ were all dynamically decoupled using the QDD scheme with an inner loop of 2 $Y$ gates and an outer loop of 2 $X$ gates. i.e. a sequence $Y$--$Y$--$X$--$Y$--$Y$--$X$--$Y$--$Y$.
    \item $q_0$, $q_1$, and $q_2$ were initialised into a $\ket{+}$ state and left idle. 
    \item $q_3$ was initialised into a $\ket{+}$ state, followed by a wait time of 320 ns, and then $q_1$ also initialised into a $\ket{+}$ state and left idle.
\end{enumerate} 

To account for the effects of measurement errors on all process tensor estimates, we first performed {GST} on the system qubit using the \texttt{pyGSTi} package~\cite{pygsti}. This was used to obtain a high quality estimate of the {POVM} $\{\ket{+}\!\bra{+}, \ket{-}\!\bra{-}, \ket{i+}\!\bra{i+},\ket{i-}\!\bra{i-}, \ket{0}\!\bra{0},\ket{1}\!\bra{1}\}$, which was then used in the {MLE} fit for each process tensor.
\end{document}